\documentclass[useAMS,usenatbib]{mn2e}

\usepackage[dvips]{graphicx}

\usepackage{lscape}
\usepackage{natbib}
\usepackage{color, ulem}

\title[Spectrum Analysis of Bright {\it Kepler} late B- to early F- Stars]{Spectrum Analysis of Bright {\it Kepler} late B- to early F- Stars\thanks{Based
on observations with the 2-m Alfred-Jensch-Telescope of the
Th\"uringer Landessternwarte Tautenburg}}
\author[A. Tkachenko et al.]{A.~Tkachenko$^{1}$\thanks{Postdoctoral Fellow of the Fund for Scientific Research (FWO), Flanders,
Belgium}, H.~Lehmann$^2$, B.~Smalley$^3$, and K.~Uytterhoeven$^{4,5}$\\
$^1$Instituut voor Sterrenkunde, K.U. Leuven, Celestijnenlaan 200D, B-3001 Leuven, Belgium\\
$^2$Th\"{u}ringer Landessternwarte Tautenburg, 07778 Tautenburg,
Germany\\
$^3$Astrophysics Group, Keele University, Staffordshire, ST5 5BG,
United Kingdom\\
$^4$ Instituto de Astrof\'{\i}sica de Canarias (IAC), Calle Via Lactea s/n, 38205 La Laguna, Tenerife, Spain\\
$^5$ Dept. Astrof\'{\i}sica, Universidad de La Laguna (ULL),
Tenerife, Spain}
\date{Received date; accepted date}
\pagerange{\pageref{firstpage}--\pageref{lastpage}} \pubyear{2011}

\def\LaTeX{L\kern-.36em\raise.3ex\hbox{a}\kern-.15em
    T\kern-.1667em\lower.7ex\hbox{E}\kern-.125emX}

\newcommand{\GD}{$\gamma$~Dor}
\newcommand{\DSct}{$\delta$~Sct}

\newcommand{\vsini}{$v\sin{i}$}
\newcommand{\cd}{c\,d$^{-1}$}
\newcommand{\te}{$T_{\rm eff}$}
\newcommand{\logg}{$\log{g}$}
\newcommand{\kms}{km\,s$^{-1}$}

\begin{document}

\label{firstpage}

\maketitle

\begin{abstract}
The $Kepler$ satellite mission was designed to search for transiting
exoplanets and delivers single band-pass light curves of a huge
number of stars observed in the Cygnus-Lyra region. At the same
time, it opens a new window for asteroseismology. In order to
accomplish one of the required preconditions for the asteroseismic
modelling of the stars, namely knowledge of their precise
fundamental parameters, ground-based spectroscopic and/or
photometric follow-up observations are needed. We aim to derive
fundamental parameters and individual abundances for a sample of 18
\GD/\DSct\ and 8 SPB/$\beta$\,Cep candidate stars in the $Kepler$
satellite field of view. We use the spectral synthesis method to
model newly obtained, high-resolution spectra of 26 stars in order
to derive their fundamental parameters like \te, \logg, \vsini,
$\xi$, [M/H], and individual abundances with high accuracy. The
stars are then placed into the log(\te)--log(g) diagram and the
obtained spectroscopic classification is compared to the existing
photometric one. For most A- and F-type stars, the derived \te\
values agree within the measurement errors with the values given in
the Kepler Input Catalog (KIC). For hot stars, the KIC temperatures
appear to be systematically underestimated, in agreement with
previous findings. We also find that the temperatures derived from
our spectra agree reasonably well with those derived from the SED
fitting. According to their position in the log(\te)--log(g)
diagram, two stars are expected \GD\ stars, four stars are expected
\DSct\ stars, and four stars are possibly \DSct\ stars at the blue
edge of the instability strip. Two stars are confirmed SPB
variables, and one star falls into the SPB instability region but
its parameters might be biased by binarity. Two of the four stars
that fall into the \DSct\ instability region show \GD-type
oscillation in their light curves implying that \GD-like
oscillations are much more common among the \DSct\ stars than is
theoretically expected. Moreover, one of the stars located at the
hot border of the \DSct\ instability strip is classified as
\DSct-\GD\ hybrid pulsator from its light curve analysis. Given that
these findings are fully consistent with recent investigations, we
conclude that a revision of the \GD\ and \DSct\ instability strips
is essential.
\end{abstract}


\section{Introduction}

Though the primary goal of actual space missions like $CoRoT$
\citep[Convection Rotation and Planetary Transits,][]{Auvergne2009}
and $Kepler$ \citep{Gilliland2010} is to search for transiting
exoplanet systems, the almost uninterrupted time series of
high-quality data led to the discovery of a huge number of pulsating
stars. This opened up a new era in asteroseismology, the study of
stellar interiors via the interpretation of pulsation patterns
observed at  the surfaces of the stars.

The amount of data delivered by these space missions is huge
implying the need in establishing methods of automatic
classification of the stars. In the case of $Kepler$ data, such
methods are usually based on the (single band-pass) light curves
morphology and/or interpretation of the corresponding Fourier
spectrum. Quite often, this leads to simultaneous assignment of the
same object to different classes of variable stars. The photometric
signal of e.g., an ellipsoidal variable can easily be misinterpreted
by rotational modulation due to stellar surface inhomogeneities or
by low-frequency stellar pulsations. Moreover, there are classes of
pulsating stars like the Gamma Doradus and SPB stars showing the
same type of variability in their light curves but residing at
different locations in the Hertzsprung-Russell diagram. The way to
discriminate between them is to derive their effective temperatures
and \logg. This is one reason why ground-based follow-up
spectroscopic and/or multi-colour photometric observations are
essential (see the ground-based follow-up campaign for Kepler
asteroseismic targets as described by Uytterhoeven et al. 2010a,b).
Moreover, high-resolution, high signal-to-noise (S/N) spectroscopic
observations allow to unreveal the nature of binary and rotationally
modulated stars by observing systematic Doppler shifts of all lines
in the spectrum or ``moving bumps'' across the line profiles of
certain chemical elements, respectively. The evaluated from
ground-based data atmospheric parameters like effective temperature
\te, surface gravity \logg, and metallicity [M/H] can further be
used for an in depth asteroseismic modelling of stars in combination
with high quality photometric data gathered by the satellites.

In this paper, we focus on SPB/$\beta$~Cep and \GD/\DSct\ candidate
stars in the $Kepler$ field of view. The term ``Slowly Pulsating B
stars'' (SPB) was introduced by \citet{Waelkens1991} who detected
multiperiodic brightness and colour variations for seven stars with
spectral types between B3 and B9. These stars have masses between 3
and 7~M$_{\odot}$. The observed photometric and radial velocity (RV)
variations  are interpreted in terms of low degree $l$, high radial
order $n$ gravity mode pulsations characterized by intrinsic periods
roughly between 0.8 and 3 days \citep[e.g.,][]{DeCat2004,Aerts2010}.
The theoretical instability strip of SPBs overlaps with the
instability region of $\beta$~Cepheii ($\beta$~Cep) pulsators, which
have higher masses (between 8 and 18~M$_{\odot}$) and are typically
hotter than SPB variables. $\beta$~Cep stars pulsate in low radial
order pressure- (p-) and gravity- (g-) modes with periods between 2
and 8 hours \citep{Aerts2010}.

\GD\ and \DSct-type stars are the two other classes of variable
pulsating stars where the theoretical instability strips overlap.
Similar to SPB stars, \GD\ stars pulsate in low degree, high order
g-modes with periods between 0.5 and 3 days \citep{Kaye1999}. It is
difficult to discriminate between \GD\ and SPB type pulsators based
on the light curve morphology and Fourier spectrum only without
having information about the temperature of the star. The observed
variability of the \DSct\ stars is understood in terms of low-order
p-modes with periods between 18 min and 8 h \citep{Aerts2010}. The
fact that \GD\ and \DSct\ instability regions overlap suggests that
{\it hybrid pulsators} showing both pulsation characteristics, i.e,
high-order g-modes and lower-order p- and g-modes, must exist.

In this paper, we present the results of the spectroscopic analysis
of 8 SPB/$\beta$~Cep and 18 \GD/\DSct\ candidate  stars in the
$Kepler$ field of view. After deriving the fundamental parameters of
our sample stars, we classify them according to the expected type of
variability and compare the results to the classification expected
from the $Kepler$ light curve analysis. For every star in the
sample, we additionally check the spectra for RV and line profile
variation (LPV) to unreveal possible binary nature of the star. The
observational material and the data reduction procedure are
described in Sect.~\ref{Section: Observations}. The method and the
results of spectrum analysis are presented in Sect.~\ref{Section:
method} and \ref{Section: results}, respectively. We discuss the
results obtained for late B- early A-type stars in
Sect.~\ref{Section: Late B- early A-type stars} and for intermediate
A- to early F-type stars in Sect.~\ref{Section: Intermediate A-
early F-type stars}. In Sect.~\ref{Section: Comparison with the
KIC}, we compare our derived fundamental parameters with the KIC
values, the overall conclusions are presented in Sect.~\ref{Section:
Conclusions}.

\section{Observations}\label{Section: Observations}

\begin{table}
\tabcolsep 2.1mm\caption{\small Journal of observations. All spectra have
been taken in 2011. $N$ gives the number of acquired spectra, $V$ the
visual magnitude, and SpT the spectral type as is indicated in the
SIMBAD database.}
\begin{tabular}{rlcrl}
\hline \multicolumn{1}{c}{KIC\rule{0pt}{9pt}} &
\multicolumn{1}{c}{Designation} & $N$
& \multicolumn{1}{c}{$V$} & \multicolumn{1}{c}{SpT}\\
\hline
02571868\rule{0pt}{9pt} & HD\,\,\,\,182271 & 3 & 8.7 & A0\\
02859567 & HD\,\,\,\,184217 & 2 & 8.3 & A0\\
02987660 & HD\,\,\,\,182634 & 3 & 8.0 & A3\\
03629496 & HD\,\,\,\,177877 & 2 & 8.3 & A0\\
04180199 & HD\,\,\,\,225718 & 9 & 10.1 & ---\\
04989900 & HD\,\,\,\,175841 & 2 & 6.9 & A2\\
05356349 & HD\,\,\,\,181680 & 2 & 8.1 & A0\\
05437206 & HD\,\,\,\,179936 & 1 & 8.4 & A2\\
06668729 & HD\,\,\,\,175536 & 1 & 8.6 & A2\\
07304385 & TYC\,3145-901-1 & 6 & 10.1 & ---\\
07827131 & HD\,\,\,\,184695 & 2 & 8.0 & A2\\
07974841 & TYC\,3148-1470-1 & 2 & 8.2 & ---\\
08018827 & HD\,\,\,\,179817 & 2 & 8.1 & B9\\
08324268 & HD\,\,\,\,189160 & 4 & 8.0 & A0p\\
08351193 & HD\,\,\,\,177152 & 4 & 7.6 & B9\\
08489712 & HD\,\,\,\,181598 & 2 & 8.6 & A0\\
08915335 & HD\,\,\,\,190566 & 6 & 9.6 & A2\\
09291618 & BD\,+452961 & 6 & 9.7 & A5\\
09351622 & BD\,+452955 & 4 & 9.1 & F0\\
10096499 & HD\,\,\,\,189013 & 2 & 6.9 & A2\\
10537907 & BD\,+472856 & 7 & 9.9 & F0\\
10974032 & HD\,\,\,\,182828 & 2 & 8.4 & A0\\
11572666 & TYC\,3565-1155-1 & 5 & 9.9 & ---\\
11874676 & BD\,+493106 & 7 & 10.1 & A5\\
12153021 & HD\,\,\,\,179617 & 3 & 8.7 & A2\\
12217324 & HD\,\,\,\,186774 & 2 & 8.3 & A0\\
\hline
\end{tabular}
\label{Table:observations}
\end{table}

We base our analysis on high-resolution, high S/N spectra taken with
the Coud\'{e}-Echelle spectrograph attached to the 2-m telescope of
the Th\"{u}ringer Landessternwarte Tautenburg. The spectra have a
resolution of 32000 and cover the wavelength range from 4720 to
7400~\AA. Table~\ref{Table:observations} represents the journal of
observations and gives the Kepler Input Catalog (KIC) number, an
alternative designation, the number of obtained spectra, the visual
magnitude, and the spectral type as is indicated in the SIMBAD
database. The number of acquired spectra is different for different
stars since we aimed to reach a S/N of about 100 for the mean,
averaged spectrum of each object.

\begin{table*}
\tabcolsep 1.5mm\center\caption{\small Fundamental stellar
parameters. The values labeled with ``K'' are taken from the KIC and
given for comparison. Metallicity values labeled with ``(Fe)'' refer
to the derived Fe abundance.}
\begin{tabular}{lllrllrllll}
\hline \multicolumn{1}{c}{KIC\rule{0pt}{9pt}} &
$T_{\rm{eff}}^{\rm{K}}(K)$ & \logg$^{\rm{K}}$ &
\multicolumn{1}{c}{$[M/H]^{\rm{K}}$} & $T_{\rm{eff}}(K)$ &
\multicolumn{1}{c}{\logg} & \multicolumn{1}{c}{$[M/H]$}
& \vsini\,(km\,s$^{-1}$) & $\xi$\,(km\,s$^{-1}$) & \multicolumn{1}{c}{SpT$^{\rm{K}}$} & \multicolumn{1}{c}{SpT}\\
\hline
02571868\rule{0pt}{11pt} & 7930 & 3.56 & --0.21 & 7880$^{+70}_{-70}$ & 3.42$^{+0.13}_{-0.13}$ & --0.22$^{+0.12}_{-0.12}$ & 205.0$^{+12.0}_{-12.0}$ & 2.80$^{+0.40}_{-0.40}$ & A6 IV-III  & A6 IV-III\vspace{1.5mm}\\
02859567 & 9418 & 4.15 & --0.05 & 9970$^{+340}_{-340}$ & 3.87$^{+0.11}_{-0.11}$ & --0.57$^{+0.20}_{-0.20}$ & 200.0$^{+25.0}_{-25.0}$ & 2.0 & A0.5 V  & B9.5 IV-V\vspace{1.5mm}\\
02987660$^{1)}$ & 7305 & 3.59 & --0.01 & 7525$^{+75}_{-75}$ & 3.46$^{+0.29}_{-0.29}$ & --0.27$^{+0.14}_{-0.14}$ & 140.0$^{+10.0}_{-10.0}$ & 2.95$^{+0.40}_{-0.40}$ & A9 IV-III  & A8 IV-III\vspace{1.5mm}\\
03629496 & 9796 & 4.50 & +0.43 & 11\,320$^{+210}_{-210}$ &
3.75$^{+0.10}_{-0.10}$ & --0.43$^{+0.20}_{-0.20}$ &
160.0$^{+19.0}_{-19.0}$ & 2.0 & B9.5 V & B8.5 IV\vspace{1.5mm}\\
04180199$^{2)}$ & 7220 & 3.91 & --0.15 & 7390$^{+80}_{-80}$ & 4.05$^{+0.32}_{-0.32}$ & --0.56$^{+0.23}_{-0.23}$ & 180.0$^{+27.0}_{-27.0}$ & 0.95$^{+0.75}_{-0.75}$ & A9.5 IV-V  & A9 IV-V\vspace{1.5mm}\\
04989900 & 7900 & 3.51 & --1.87 & 8400$^{+150}_{-150}$ & 3.08$^{+0.11}_{-0.11}$ & --0.23$^{+0.22}_{-0.22}$ & 191.0$^{+15.0}_{-15.0}$ & 2.33$^{+0.65}_{-0.65}$ & A6 IV-III  & A4 III-II\vspace{1.5mm}\\
05356349 & 8295 & 3.91 & --0.06 & 8820$^{+180}_{-180}$ & 3.51$^{+0.11}_{-0.11}$ & --0.55$^{+0.27}_{-0.27}$ & 197.0$^{+24.0}_{-24.0}$ & 2.06$^{+0.74}_{-0.74}$ & A5 IV-V  & A2.5 IV-III\vspace{1.5mm}\\
05437206 & 7710 & 3.67 & --0.03 & 7870$^{+100}_{-100}$ & 3.10$^{+0.13}_{-0.13}$ & --0.24$^{+0.14}_{-0.14}$ & 125.5$^{+7.5}_{-7.5}$ & 2.75$^{+0.45}_{-0.45}$ & A7 IV  & A6 III\vspace{1.5mm}\\
06668729 & 7770 & 3.49 & --0.16 & 7800$^{+75}_{-75}$ & 3.49$^{+0.25}_{-0.25}$ & --0.45$^{+0.15}_{-0.15}$ & 128.0$^{+9.0}_{-9.0}$ & 2.33$^{+0.45}_{-0.45}$ & A6.5 IV-III  & A6.5 IV-III\vspace{1.5mm}\\
07304385$^{1)}$ & 6890 & 3.60 & --0.06 & 7020$^{+70}_{-70}$ & 3.65$^{+0.25}_{-0.25}$ & --0.32$^{+0.10}_{-0.10}$ & 64.5$^{+2.8}_{-2.8}$ & 2.45$^{+0.25}_{-0.25}$ & F2 IV  & F1 IV\vspace{1.5mm}\\
07827131 & 8285 & 3.49 & --0.19 & 8015$^{+120}_{-120}$ & 2.79$^{+0.10}_{-0.10}$ & --1.15(Fe) & 228.0$^{+26.0}_{-26.0}$ & 4.85$^{+0.85}_{-0.85}$ & A4.5 IV-III  & A6 II-III\vspace{1.5mm}\\
07974841 & 8930 & 3.82 & --0.15 & 10\,650$^{+285}_{-285}$ & 3.87$^{+0.14}_{-0.14}$ & +0.00$^{+0.13}_{-0.13}$ & 33.0$^{+5.0}_{-5.0}$ & 2.0 & A2 IV-V & B9 IV-V\vspace{1.5mm}\\
08018827 & 9188 & 3.65 & --0.14 & 10\,945$^{+350}_{-350}$ & 3.98$^{+0.16}_{-0.16}$ & --0.44$^{+0.25}_{-0.25}$ & 243.0$^{+32.0}_{-32.0}$ & 2.0 & A1.5 IV-III & B8.5 V-IV\vspace{1.5mm}\\
08324268$^{1)}$ & 9045 & 4.32 & +0.27 & 11\,370$^{+440}_{-440}$
& 3.35$^{+0.20}_{-0.20}$ & +0.65$^{+0.13}_{-0.13}$ & 31.0$^{+4.0}_{-4.0}$ & 2.0 & A1.5 V  & B8.5 III\vspace{1.5mm}\\
08351193 & 8467 & 3.99 & --0.14 & 9980$^{+250}_{-250}$ &
3.80$^{+0.15}_{-0.15}$ & --2.35(Fe) & 180.0$^{+29.0}_{-29.0}$ & 2.0 & A4 IV-V & A0 IV-V\vspace{1.5mm}\\
08489712 & 8350 & 3.52 & --0.33 & 8270$^{+150}_{-150}$ & 2.90$^{+0.15}_{-0.15}$ & --0.60$^{+0.25}_{-0.25}$ & 119.0$^{+14.0}_{-14.0}$ & 1.25$^{+0.70}_{-0.70}$ & A4 IV-III  & A4.5 III-II\vspace{1.5mm}\\
08915335 & 7770 & 3.48 & --0.13 & 8000$^{+100}_{-100}$ & 3.15$^{+0.11}_{-0.11}$ & --0.10$^{+0.20}_{-0.20}$ & 200.0$^{+17.0}_{-17.0}$ & 1.58$^{+0.40}_{-0.40}$ & A6.5 IV-III  & A5.5 III\vspace{1.5mm}\\
09291618 & 7610 & 3.61 & --0.08 & 7530$^{+75}_{-75}$ & 3.56$^{+0.40}_{-0.40}$ & --0.36$^{+0.20}_{-0.20}$ & 177.0$^{+18.0}_{-18.0}$ & 1.10$^{+0.60}_{-0.60}$ & A7.5 IV-III  & A8 IV-III\vspace{1.5mm}\\
09351622$^{1)}$ & 7450 & 3.53 & --0.22 & 7515$^{+70}_{-70}$ & 3.17$^{+0.30}_{-0.30}$ & --0.24$^{+0.12}_{-0.12}$ & 78.0$^{+4.5}_{-4.5}$ & 2.90$^{+0.35}_{-0.35}$ & A8 IV-III  & A7.5 III\vspace{1.5mm}\\
10096499 & 7780 & 4.13 & --0.02 & 7960$^{+85}_{-85}$ & 3.27$^{+0.12}_{-0.12}$ & --0.60$^{+0.13}_{-0.13}$ & 89.5$^{+5.5}_{-5.5}$ & 2.78$^{+0.40}_{-0.40}$ & A7 V-IV  & A5.5 III-IV\vspace{1.5mm}\\
10537907$^{1)}$ & 7500 & 3.45 & --0.27 & 7400$^{+70}_{-70}$ & 3.51$^{+0.30}_{-0.30}$ & --0.45$^{+0.14}_{-0.14}$ & 112.0$^{+8.0}_{-8.0}$ & 2.80$^{+0.40}_{-0.40}$ & A8 IV-III  & A8.5 IV-III\vspace{1.5mm}\\
10974032 & 9038 & 3.70 & --0.33 & 9750$^{+370}_{-370}$ & 3.75$^{+0.20}_{-0.20}$ & --0.80(Fe) & 270.0$^{+32.0}_{-32.0}$ & 2.0 & A2 IV & A0 IV\vspace{1.5mm}\\
11572666$^{2)}$ & 7040 & 3.49 & --0.62 & 7265$^{+85}_{-85}$ & 4.10$^{+0.45}_{-0.45}$ & --1.00(Fe) & 180.5$^{+31.0}_{-31.0}$ & 1.90$^{+0.85}_{-0.85}$ & F1 IV-III  & A9.5 V-IV\vspace{1.5mm}\\
11874676$^{2)}$ & 8220 & 4.00 & --0.14 & 7885$^{+80}_{-80}$ & 3.57$^{+0.25}_{-0.25}$ & --1.00(Fe) & 203.5$^{+27.0}_{-27.0}$ & 2.97$^{+0.75}_{-0.75}$ & A5 IV-V  & A6 IV-III\vspace{1.5mm}\\
12153021 & 9041 & 3.90 & --0.11 & 9010$^{+80}_{-80}$ & 3.50$^{+0.10}_{-0.10}$ & --0.05$^{+0.15}_{-0.15}$ & 18.0$^{+2.7}_{-2.7}$ & 2.02$^{+0.65}_{-0.65}$ & A2 IV-V & A2 IV-III\vspace{1.5mm}\\
12217324 & 10\,434 & 3.93 & --0.09 & 10\,380$^{+270}_{-270}$ & 3.75$^{+0.15}_{-0.15}$ & +0.22$^{+0.14}_{-0.14}$ & 19.0$^{+3.5}_{-3.5}$ & 2.0 & B9 V-IV & B9.5 IV\vspace{1.5mm}\\
\hline \multicolumn{11}{l}{$^{1)}$ LPV detected; $^{2)}$ Stars with
composite spectra}\rule{0pt}{11pt}
\end{tabular}
\label{Table:FundamentalParameters}
\end{table*}

The data were reduced using standard ESO-MIDAS packages. The data
reduction included bias and stray-light subtraction, cosmic rays
filtering, flat fielding using a halogen lamp, wavelength
calibration using a ThAr lamp, and normalisation to the local
continuum. All spectra were additionally corrected in wavelength for
individual instrumental shifts by using a large number of telluric
O$_2$ lines. The cross-correlation technique was used to measure the
RVs from the single spectra so that the single spectra finally could
be shifted and co-added to build the mean, high S/N averaged
spectrum of each star.

We use publicly available both long- (time-resolution $\sim$~30
min.) and short-cadence (time-resolution $\sim$~1 min.) {\it Kepler}
data to make an additional check for binarity, stellar activity and
pulsations for each star in the sample. The {\it Kepler} data are
released in quarters, i.e. periods between two spacecraft rolls. For
this study, we use the data from quarters Q0-Q6 (May 2009 -
September 2010) where available.

\begin{figure}
\includegraphics[scale=0.9,clip=]{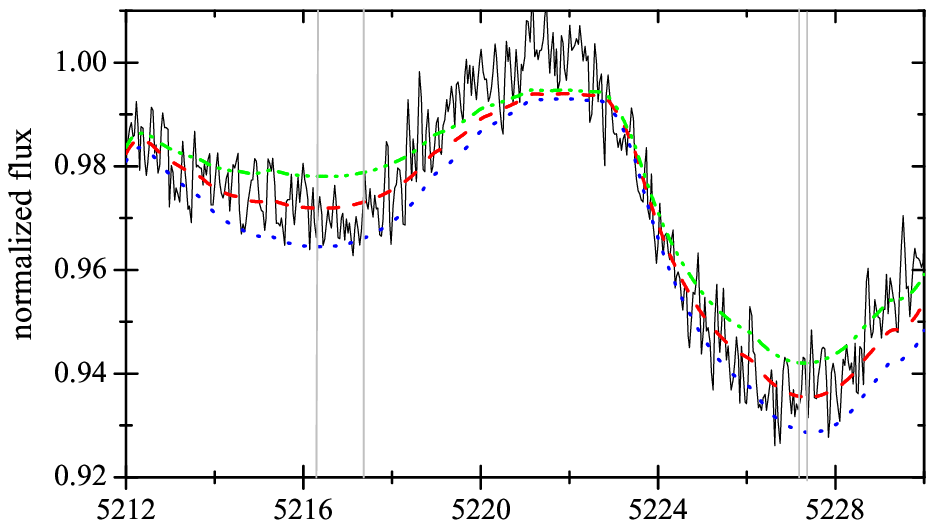}\vspace{2mm}
\includegraphics[scale=0.9,clip=]{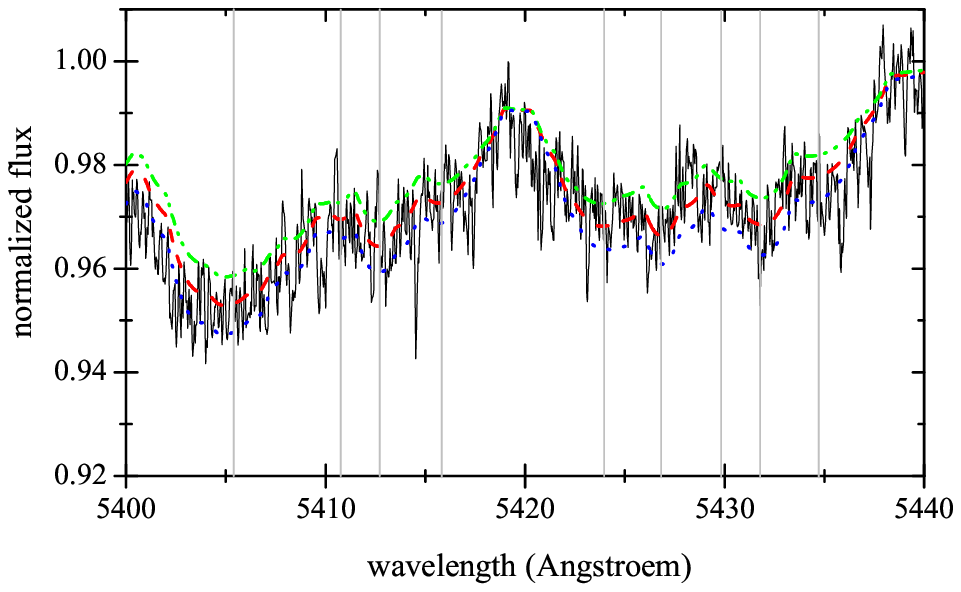}
\caption{{\small Fit of a part of the observed spectrum of
KIC\,02571868 (black, solid line) by synthetic spectra computed from
our optimized parameters (red, dashed line) and assuming Fe
abundance to vary within the quoted error bars of $\pm$0.15~dex
(blue, dotted and green, dash dot dotted lines, respectively).
Vertical light gray lines indicate positions of (strongest) Fe
lines.}} \label{K02571868}
\end{figure}

\section{Method}\label{Section: method}

Our code GSSP \citep{Tkachenko2012} finds the optimum values of \te,
\logg, $\xi$, $[M/H]$, and \vsini\ from the minimum in $\chi^2$
obtained from a comparison of the observed spectrum with the
synthetic ones computed from all possible combinations of the above
mentioned parameters. The errors of measurement (1$\sigma$
confidence level) are calculated from the $\chi^2$ statistics using
the projections of the hypersurface of the $\chi^2$ from all grid
points of all parameters onto the parameter in question. In this
way, the estimated error bars include any possible model-inherent
correlations between the parameters. Possible imperfection of the
model like incorrect atomic data, non-LTE effects, or continuum
normalization are not taken into account, of course.
\citet{Fossati2007, Fossati2008} state that the continuum
normalization is a source of abundance uncertainty that can raise
from about 0.1 to 0.2~dex for large \vsini. We corrected the
observed spectra during the analysis for large-scale imperfections
of the continua by adjusting them to the model continua. The errors
of the fit become a bit larger in this way, due to the inclusion of
additional free parameters. Thus we believe that the latter value
mentioned by \citet{Fossati2007, Fossati2008} is an upper limit
given that the abundance uncertainty due to small-scale continuum
imperfections should decrease with increasing number of analysed
spectral lines.

In a recent study by \citet{Molenda-Zakowicz2013} the use of
different methods and codes to derive atmospheric parameters for F,
G, K, and M-type stars is compared, and lead the authors to conclude
that the realistic accuracy in the determination of atmospheric
parameters for these types of stars is $\pm$~150~K in \te,
$\pm$~0.15~dex in [Fe/H], and 0.3~dex in \logg, even though error
calculations for individual programs might result in smaller errors.
Hence, we are aware of a possible underestimation of errors in
Table~\ref{Table:FundamentalParameters}.

In order to check how reliable our error estimates on elemental
abundances are, we made an additional test represented in
Figure~\ref{K02571868}. We have chosen KIC\,02571868, the star with
large value of \vsini\ of $\sim$200~\kms, as its broad and rather
shallow lines should have a lower sensitivity to the elemental
abundance changes than, e.g., the lines of slowly rotating star
KIC\,12217324. We would thus expect error bars to be larger for the
star with broad lines, at least for the chemical elements
represented by sufficient number of individual lines in the stellar
spectrum (e.g., Fe). Figure~\ref{K02571868} compares a part of the
observed spectrum of KIC\,02571868 with synthetic spectra computed
assuming different Fe abundance that varies within the quoted error
bars of $\pm$0.15~dex (see Table~\ref{Table:IndividualAbundances}).
These rather small changes in Fe abundance cause quite clear
deviations in the synthetic spectra and hence the difference in the
quality of the fit. The corresponding values of $\chi^2$ deviate
from the optimal fit value by slightly more than 1$\sigma$
confidence level which is assumed to represent the errors of
measurement as described above. The fact that the abundances of
elements like Ti, Ca, etc. are derived for slowly rotating star
KIC\,12217324 with lower precision than it is actually for the fast
rotator KIC\,02571868, is due to significant difference in
atmospheric parameters, and the temperature in particular. The above
mentioned elements are represented by a few, rather weak lines in
the spectrum of KIC\,12217324, which obviously makes estimation of
their abundances more challenging in this case and facilitates an
increase of the corresponding error bars.

\begin{figure}
\includegraphics[scale=0.9,clip=]{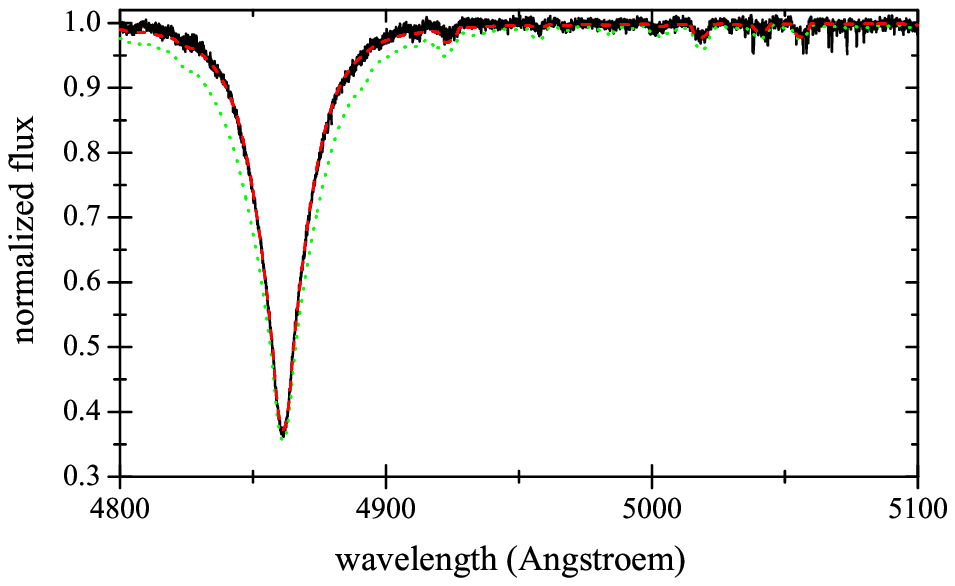}
\includegraphics[scale=0.9,clip=]{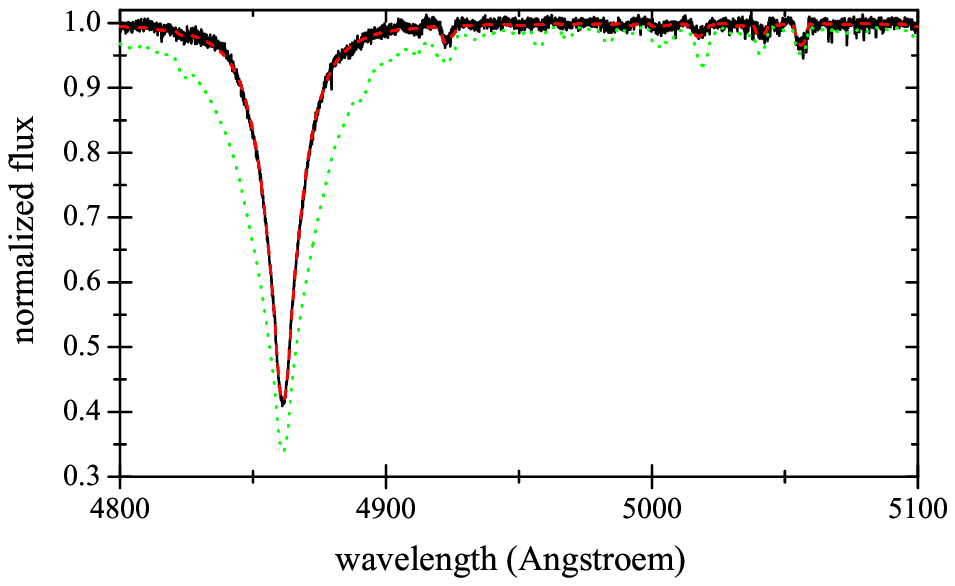}
\caption{{\small Fit of the observed (black, solid line) by
synthetic spectra computed from our optimized parameters (red,
dashed line) and from the values given in the KIC (green, dotted
line). From top to bottom: KIC\,02859567 and 03629496}}
\label{K02859567}
\end{figure}

A detailed description of the method and its application to the
spectra of $Kepler$ $\beta$\,Cep and SPB candidate stars as well as
$\delta$\,Sct and $\gamma$\,Dor candidate stars are given in
\citet{Lehmann2011} and \citet{Tkachenko2012}, respectively.

For the calculation of synthetic spectra, we use the LTE-based
code SynthV \citep{Tsymbal1996} which allows to compute the
spectra based on individual elemental abundances. The code uses
pre-calculated atmosphere models which have been computed with the
most recent, parallelised version of the LLmodels program
\citep{Shulyak2004}. Both programs make use of the VALD database
\citep{Kupka2000} for a pre-selection of atomic spectral lines.
The main limitation of the LLmodels code is that the models are
well suitable for early and intermediate spectral type stars but
not for very hot and cool stars where non-LTE effects or
absorption in molecular bands may become relevant, respectively.

\section{Results}\label{Section: results}

Table~\ref{Table:FundamentalParameters} summarizes the results of
spectrum analysis for all stars of our sample. The first four
columns of the table represent correspondingly the KIC-number of the
star, and effective temperature $T_{\rm{eff}}^{\rm{K}}$, surface
gravity $\log{g}^{\rm{K}}$, and metallicity $[M/H]^{\rm{K}}$ as is
indicated in the KIC. The five following columns list the stellar
parameters derived from our spectra, while the last two columns
represent the spectral types as estimated from \te\ and \logg\ given
in the KIC and determined in this work, respectively. In both cases,
the spectral types and the luminosity classes have been derived
using an interpolation in the tables published by
\citet{Schmidt-Kaler1982}. Metallicity values labeled with ``(Fe)''
refer to the derived Fe abundance, the corresponding measurement
errors are given in Table~\ref{Table:IndividualAbundances}. For the
eight hottest stars of the sample (late B- to early A-type stars),
the micro-turbulent velocity was fixed to a standard value of
2~km\,s$^{-1}$ because of the strong correlation between $\xi$ and
$T_{\rm{eff}}$ for higher temperatures (see Figure~2 in
\citet{Lehmann2011}). In practice, this means that the errors in
$\xi$ raise up to about 1.5~km\,s$^{-1}$ for stars with
$T_{\rm{eff}} > 10\,000\,{\rm K}$ and to about 3--4~km\,s$^{-1}$ for
stars hotter than 15\,000~K. The results thus appear to be almost
insensitive to the microturbulent velocity in this case.

Table~\ref{Table:IndividualAbundances} lists the elemental
abundances derived for each target star. The metallicity given in
the second column of the table refers to the initially derived
chemical composition and was used as initial guess for the
determination of the individual abundances. The latter are given
relative to solar values, i.e. negative/positive values refer to an
under-/overabundance of the corresponding element compared to the
solar composition. We assume the chemical composition of the Sun
given by \citet{Grevesse2007}, corresponding values are listed in
the header of Table~\ref{Table:IndividualAbundances} below the
element designation. For five stars we have reached  the lower
metallicity limit of --0.8~dex in our grid of atmosphere models
(KIC\,07827131, 08351193, 10974032, 11572666 and 11874676) and give
the derived Fe abundance instead. In all other cases, the derived Fe
abundance matches the derived metallicity within the quoted error
bars.

Similar to the results reported in two of our previous papers
\citep{Lehmann2011,Tkachenko2012}, we find that the temperatures
listed in the KIC are in general underestimated for the hotter
stars. In the following, we discuss the results on individual stars
in more detail.

\subsection{Late B- early A-type stars}\label{Section: Late B- early A-type
stars}

Two of eight stars, {\it KIC\,08351193} and {\it 10974032}, are
found to be metal poor with metallicities below the lower limit of
--0.8~dex in our grid of atmosphere models. Both stars show nearly
solar He abundance and overabundance of Mg compared to the derived
Fe content, while the spectrum of KIC\,08351193 additionally
exhibits strong enhancement of Si. Both are fast rotators and hotter
than is expected from the KIC. The surface gravity is consistent
with the one listed in the KIC in both cases. \citet{Balona2011b}
reports the effective temperatures of 10\,210~K and
11\,000$\pm$400~K for KIC\,08351193, determined from Str\"{o}mgren
photometry and from spectral energy distribution (SED) fitting to
the combined SDSS, 2MASS, and Str\"{o}mgren colours, accordingly.
The first value agrees within the error bars with the one we
determine spectroscopically in this study. The two spectra we
obtained for KIC\,10974032 is not enough to conclude about the
presence of line profile variation (LPV). The light curve of this
star is irregular, typical of that of an active star. The four
spectra of KIC\,08351193 do not show LPV either, though the star is
classified as rotationally modulated by
\citet{Balona2011b}.

{\it KIC\,02859567} and {\it 03629496:} two fast rotating late
B-type stars showing large deviations of the derived parameters from
those listed in the KIC. Both stars show significant underabundances
of Fe compared to the solar composition though consistent with the
derived metallicity in both cases. Both show about solar He content
and slight overabundances of Si. The spectrum of KIC\,02859567
additionally exhibits an excess in Mg. Figure~\ref{K02859567}
compares the observations with two synthetic spectra computed from
our optimised and the KIC parameters in a small wavelength region
around H$_{\beta}$. The big difference in the quality of the fit is
primarily explained by the difference in the assumed temperatures
and confirms the conclusion made by\citet{Molenda-Zakowicz2010} that
$T_{\rm eff}$ given in the KIC is too low for stars hotter than
about 7\,000~K. The deviation in general becomes larger the hotter
the stars. No signatures of binarity nor stellar activity could be
found in the light curves of both stars. KIC\,03629496 is classified
by \citet{Balona2011a} as a beating star with a dominant
contribution in the frequency range  between 2 and 4~\cd.
KIC\,02859567 in turn exhibits very low-amplitude photometric
variability in the same frequency regime making it a SPB
candidate.

{\it KIC\,07974841} and {\it 08018827:} Both stars show clear
variability in their light curves. KIC\,07974841 is found in the
Washington Double Star Catalog as WDS\,19466+4346. The primary and
secondary components are assumed to be of 8.24 and 11.32 mag,
correspondingly; the magnitude of the primary is consistent with the
visual magnitude found in the SIMBAD database. The {\it Kepler}
light curve of the star is too irregular to be explained by
ellipsoidal effects occurring in close binary systems and is rather
due to rotational modulation, stellar pulsations or a combination of
both. \citet{Balona2011b} classify the star as rotationally
modulated SPB variable, where the rotation signal dominates the
pulsation one. Our two spectra do not show any variability that
could be attributed to the binary nature of the star, neither global
Doppler shifts of spectral lines nor any signature of the second
star in the residuals of spectrum fitting.\linebreak

\begin{landscape}
\begin{table}
\tabcolsep 0.9mm \caption{\small Metallicity and elemental
abundances relative to solar ones in dex. Metallicity values labeled
with ``(Fe)'' refer to the derived Fe abundance.}
\begin{tabular}{ccccccccccccccc}
\hline
KIC\rule{0pt}{9pt} & $[M/H]$ & \multicolumn{1}{c}{He} & \multicolumn{1}{c}{Fe} & \multicolumn{1}{c}{Mg} & \multicolumn{1}{c}{Si} & \multicolumn{1}{c}{Ti} & \multicolumn{1}{c}{Cr}& \multicolumn{1}{c}{O} & \multicolumn{1}{c}{Ca} & \multicolumn{1}{c}{Sc} & \multicolumn{1}{c}{Ni} & \multicolumn{1}{c}{C} & \multicolumn{1}{c}{Mn} & \multicolumn{1}{c}{Y}\\
    &            & \multicolumn{1}{c}{--1.11} & \multicolumn{1}{c}{--4.59} & \multicolumn{1}{c}{--4.51} & \multicolumn{1}{c}{--4.53} & \multicolumn{1}{c}{--7.14} & \multicolumn{1}{c}{--6.40} & \multicolumn{1}{c}{--3.38} & \multicolumn{1}{c}{--5.73} & \multicolumn{1}{c}{--8.99} & \multicolumn{1}{c}{--5.81} & \multicolumn{1}{c}{--3.65} & \multicolumn{1}{c}{--6.65} & \multicolumn{1}{c}{--9.83}\\
\hline
02571868\rule{0pt}{11pt} & --0.22$^{+0.12}_{-0.12}$ & --- & --0.25$^{+0.15}_{-0.15}$ & --0.10$^{+0.40}_{-0.40}$ & +0.20$^{+0.35}_{-0.35}$ & --0.50$^{+0.25}_{-0.25}$ & --0.05$^{+0.20}_{-0.20}$ & --0.60$^{+0.60}_{-0.60}$ & +0.00$^{+0.20}_{-0.20}$ & --0.10$^{+0.40}_{-0.40}$ & --0.45$^{+0.30}_{-0.30}$ & --0.15$^{+0.25}_{-0.25}$ & --0.65$^{+0.45}_{-0.45}$ & --0.15$^{+0.40}_{-0.40}$\vspace{2.0mm}\\
02859567 & --0.57$^{+0.20}_{-0.20}$ & +0.07$^{+0.20}_{-0.20}$ & --0.75$^{+0.25}_{-0.25}$ & --0.15$^{+0.35}_{-0.35}$ & --0.25$^{+0.35}_{-0.35}$ & --- & ---& --- & --- & --- & --- & --- & --- & ---\vspace{2.0mm}\\
02987660 & --0.27$^{+0.14}_{-0.14}$ & --- & --0.35$^{+0.16}_{-0.16}$ & +0.20$^{+0.31}_{-0.31}$ & +0.05$^{+0.40}_{-0.40}$ & --0.30$^{+0.26}_{-0.26}$ & --0.40$^{+0.25}_{-0.25}$ & --- & --0.35$^{+0.31}_{-0.31}$ & --0.25$^{+0.45}_{-0.45}$ & --0.20$^{+0.26}_{-0.26}$ & --0.05$^{+0.25}_{-0.25}$ & --0.05$^{+0.34}_{-0.34}$ & --0.40$^{+0.40}_{-0.40}$\vspace{2.0mm}\\
03629496 & --0.43$^{+0.20}_{-0.20}$ & --0.07$^{+0.20}_{-0.20}$ & --0.55$^{+0.25}_{-0.25}$ & --0.45$^{+0.35}_{-0.35}$ & --0.05$^{+0.35}_{-0.35}$ & --- & ---& --- & --- & --- & --- & --- & --- & ---\vspace{2.0mm}\\
04180199 & --0.56$^{+0.23}_{-0.23}$ & --- & --0.75$^{+0.25}_{-0.25}$ & --0.15$^{+0.35}_{-0.35}$ & --0.20$^{+0.75}_{-0.75}$ & --0.85$^{+0.50}_{-0.50}$ & --0.50$^{+0.50}_{-0.50}$ & --- & --0.35$^{+0.50}_{-0.50}$ & --0.55$^{+0.60}_{-0.60}$ & --0.60$^{+0.45}_{-0.45}$ & --0.20$^{+0.60}_{-0.60}$ & --0.35$^{+0.75}_{-0.75}$ & ---\vspace{2.0mm}\\
04989900 & --0.23$^{+0.22}_{-0.22}$ & --- & --0.20$^{+0.22}_{-0.22}$ & --0.15$^{+0.40}_{-0.40}$ & +0.10$^{+0.46}_{-0.46}$ & --0.50$^{+0.35}_{-0.35}$ & --0.30$^{+0.32}_{-0.32}$ & --0.10$^{+0.36}_{-0.36}$ & --0.45$^{+0.35}_{-0.35}$ & +0.10$^{+0.55}_{-0.55}$ & --0.05$^{+0.41}_{-0.41}$ & +0.00$^{+0.32}_{-0.32}$ & --0.20$^{+0.45}_{-0.45}$ & ---\vspace{2.0mm}\\
05356349 & --0.55$^{+0.27}_{-0.27}$ & --- & --0.70$^{+0.23}_{-0.23}$ & --0.05$^{+0.42}_{-0.42}$ & --0.20$^{+0.50}_{-0.50}$ & --1.15$^{+0.41}_{-0.41}$ & --1.05$^{+0.65}_{-0.65}$ & --0.40$^{+0.65}_{-0.65}$ & --0.60$^{+0.47}_{-0.47}$ & --- & --- & --0.70$^{+0.66}_{-0.66}$ & --- & ---\vspace{2.0mm}\\
05437206 & --0.24$^{+0.14}_{-0.14}$ & --- & --0.20$^{+0.17}_{-0.17}$ & +0.20$^{+0.41}_{-0.41}$ & +0.10$^{+0.57}_{-0.57}$ & --0.40$^{+0.26}_{-0.26}$ & --0.15$^{+0.27}_{-0.27}$ & --0.05$^{+0.40}_{-0.40}$ & --0.25$^{+0.31}_{-0.31}$ & --0.10$^{+0.35}_{-0.35}$ & --0.20$^{+0.26}_{-0.26}$ & --0.05$^{+0.26}_{-0.26}$ & --0.45$^{+0.42}_{-0.42}$ & --0.20$^{+0.45}_{-0.45}$\vspace{2.0mm}\\
06668729 & --0.45$^{+0.15}_{-0.15}$ & --- & --0.40$^{+0.17}_{-0.17}$ & +0.00$^{+0.41}_{-0.41}$ & --0.05$^{+0.56}_{-0.56}$ & --0.80$^{+0.40}_{-0.40}$ & --0.45$^{+0.36}_{-0.36}$ & --0.25$^{+0.65}_{-0.65}$ & --0.35$^{+0.35}_{-0.35}$ & --0.40$^{+0.45}_{-0.45}$ & --0.60$^{+0.41}_{-0.41}$ & --0.20$^{+0.35}_{-0.35}$ & --0.25$^{+0.50}_{-0.50}$ & --0.40$^{+0.50}_{-0.50}$\vspace{2.0mm}\\
07304385 & --0.32$^{+0.10}_{-0.10}$ & --- & --0.30$^{+0.12}_{-0.12}$ & --0.10$^{+0.21}_{-0.21}$ & --0.15$^{+0.30}_{-0.30}$ & --0.35$^{+0.22}_{-0.22}$ & --0.25$^{+0.22}_{-0.22}$ & --- & --0.20$^{+0.20}_{-0.20}$ & --0.25$^{+0.30}_{-0.30}$ & --0.30$^{+0.21}_{-0.21}$ & --0.10$^{+0.25}_{-0.25}$ & --0.20$^{+0.32}_{-0.32}$ & --0.30$^{+0.35}_{-0.35}$\vspace{2.0mm}\\
07827131 & --1.15(Fe) & --- & --1.15$^{+0.22}_{-0.22}$ & -0.80$^{+0.31}_{-0.31}$ & --0.55$^{+0.50}_{-0.50}$ & --1.85$^{+0.75}_{-0.75}$ & --1.30$^{+0.67}_{-0.67}$ & --- & --1.00$^{+0.50}_{-0.50}$ & --- & --- & --0.90$^{+0.52}_{-0.52}$ & --- & ---\vspace{2.0mm}\\
07974841 & +0.00$^{+0.13}_{-0.13}$ & --0.92$^{+0.20}_{-0.20}$ & +0.10$^{+0.20}_{-0.20}$ & --0.35$^{+0.35}_{-0.35}$ & --0.05$^{+0.35}_{-0.35}$ & +0.15$^{+0.35}_{-0.35}$ & +0.35$^{+0.35}_{-0.35}$ & --- & --- & --- & --- & --- & --- & ---\vspace{2.0mm}\\
08018827 & --0.44$^{+0.25}_{-0.25}$ & +0.06$^{+0.20}_{-0.20}$ & --0.45$^{+0.25}_{-0.25}$ & +0.55$^{+0.35}_{-0.35}$ & --0.55$^{+0.35}_{-0.35}$ & ---& --- & --- & --- & --- & --- & --- & --- & ---\vspace{2.0mm}\\
08324268 & +0.65$^{+0.13}_{-0.13}$ & \multicolumn{13}{c}{{\small{\bf abundances are not evaluated}}}\vspace{2.0mm}\\
08351193 & --2.35(Fe) & +0.05$^{+0.20}_{-0.20}$ & --2.35$^{+0.25}_{-0.25}$ & --1.40$^{+0.35}_{-0.35}$ & --1.20$^{+0.35}_{-0.35}$ & --- & --- & --- & --- & --- & --- & --- & --- & ---\vspace{2.0mm}\\
08489712 & --0.60$^{+0.25}_{-0.25}$ & --- & --0.50$^{+0.23}_{-0.23}$ & --0.30$^{+0.47}_{-0.47}$ & +0.00$^{+0.47}_{-0.47}$ & --0.90$^{+0.42}_{-0.42}$ & --0.50$^{+0.48}_{-0.48}$ & --- & --0.60$^{+0.45}_{-0.45}$ & --- & --- & --0.30$^{+0.52}_{-0.52}$ & --- & ---\vspace{2.0mm}\\
08915335 & --0.10$^{+0.20}_{-0.20}$ & --- & --0.05$^{+0.17}_{-0.17}$ & +0.35$^{+0.40}_{-0.40}$ & +0.80$^{+0.32}_{-0.32}$ & --0.40$^{+0.40}_{-0.40}$ & +0.20$^{+0.27}_{-0.27}$ & --0.30$^{+0.65}_{-0.65}$ & +0.00$^{+0.36}_{-0.36}$ & --0.35$^{+0.55}_{-0.55}$ & --0.25$^{+0.37}_{-0.37}$ & --0.20$^{+0.42}_{-0.42}$ & --0.35$^{+0.50}_{-0.50}$ & ---\vspace{2.0mm}\\
09291618 & --0.36$^{+0.20}_{-0.20}$ & --- & --0.40$^{+0.23}_{-0.23}$ & --0.35$^{+0.52}_{-0.52}$ & +0.20$^{+0.40}_{-0.40}$ & --0.70$^{+0.45}_{-0.45}$ & --0.55$^{+0.40}_{-0.40}$ & --- & --0.60$^{+0.52}_{-0.52}$ & --0.50$^{+0.65}_{-0.65}$ & --0.30$^{+0.42}_{-0.42}$ & --0.10$^{+0.35}_{-0.35}$ & --0.05$^{+0.65}_{-0.65}$ & ---\vspace{2.0mm}\\
09351622 & --0.24$^{+0.12}_{-0.12}$ & --- & --0.25$^{+0.15}_{-0.15}$ & +0.00$^{+0.30}_{-0.30}$ & +0.00$^{+0.30}_{-0.30}$ & --0.40$^{+0.20}_{-0.20}$ & --0.15$^{+0.20}_{-0.20}$ & --- & --0.15$^{+0.21}_{-0.21}$ & --0.10$^{+0.30}_{-0.30}$ & --0.25$^{+0.22}_{-0.22}$ & --0.15$^{+0.25}_{-0.25}$ & --0.15$^{+0.30}_{-0.30}$ & --0.30$^{+0.35}_{-0.35}$\vspace{2.0mm}\\
10096499 & --0.60$^{+0.13}_{-0.13}$ & --- & --0.60$^{+0.15}_{-0.15}$ & --0.10$^{+0.30}_{-0.30}$ & --0.30$^{+0.38}_{-0.38}$ & --0.80$^{+0.29}_{-0.29}$ & --0.55$^{+0.35}_{-0.35}$ & --0.30$^{+0.65}_{-0.65}$ & --0.45$^{+0.30}_{-0.30}$ & --0.65$^{+0.42}_{-0.42}$ & --0.55$^{+0.35}_{-0.35}$ & --0.45$^{+0.35}_{-0.35}$ & --0.70$^{+0.51}_{-0.51}$ & ---\vspace{2.0mm}\\
10537907 & --0.45$^{+0.14}_{-0.14}$ & --- & --0.60$^{+0.20}_{-0.20}$ & --0.20$^{+0.40}_{-0.40}$ & --0.20$^{+0.50}_{-0.50}$ & --0.50$^{+0.25}_{-0.25}$ & --0.35$^{+0.25}_{-0.25}$ & --- & --0.25$^{+0.31}_{-0.31}$ & --0.40$^{+0.50}_{-0.50}$ & --0.55$^{+0.29}_{-0.29}$ & --0.15$^{+0.35}_{-0.35}$ & --0.40$^{+0.40}_{-0.40}$ & ---\vspace{2.0mm}\\
10974032 & --0.80(Fe) & --0.04$^{+0.20}_{-0.20}$ & --0.80$^{+0.27}_{-0.27}$ & +0.15$^{+0.38}_{-0.38}$ & --0.50$^{+0.36}_{-0.36}$ & --- & --- & --- & --- & --- & --- & --- & --- & ---\vspace{2.0mm}\\
11572666 & --1.00(Fe) & --- & --1.00$^{+0.25}_{-0.25}$ & --0.40$^{+0.40}_{-0.40}$ & --0.10$^{+0.57}_{-0.57}$ & --0.95$^{+0.48}_{-0.48}$ & --1.00$^{+0.48}_{-0.48}$ & --- & --0.80$^{+0.47}_{-0.47}$ & --1.30$^{+0.65}_{-0.65}$ & --0.65$^{+0.45}_{-0.45}$ & --0.25$^{+0.45}_{-0.45}$ & --0.70$^{+0.67}_{-0.67}$ & ---\vspace{2.0mm}\\
11874676 & --1.00(Fe) & --- & --1.00$^{+0.25}_{-0.25}$ & --0.10$^{+0.47}_{-0.47}$ & --0.05$^{+0.48}_{-0.48}$ & --1.00$^{+0.50}_{-0.50}$ & --1.20$^{+0.50}_{-0.50}$ & --- & --0.65$^{+0.47}_{-0.47}$ & --1.00$^{+0.67}_{-0.67}$ & --0.75$^{+0.45}_{-0.45}$ & --0.55$^{+0.65}_{-0.65}$ & --- & ---\vspace{2.0mm}\\
12153021 & --0.05$^{+0.15}_{-0.15}$ & +0.00$^{+0.20}_{-0.20}$ & --0.10$^{+0.18}_{-0.18}$ & +0.00$^{+0.27}_{-0.27}$ & --0.10$^{+0.38}_{-0.38}$ & --0.15$^{+0.26}_{-0.26}$ & +0.05$^{+0.27}_{-0.27}$& --0.05$^{+0.38}_{-0.38}$ & --0.30$^{+0.45}_{-0.45}$ & --0.35$^{+0.45}_{-0.45}$ & --- & --- & --- & ---\vspace{2.0mm}\\
12217324 & +0.22$^{+0.14}_{-0.14}$ & +0.00$^{+0.20}_{-0.20}$ & +0.30$^{+0.15}_{-0.15}$ & --0.15$^{+0.35}_{-0.35}$ & +0.10$^{+0.35}_{-0.35}$ & +0.25$^{+0.35}_{-0.35}$ & +0.55$^{+0.35}_{-0.35}$& --0.25$^{+0.35}_{-0.35}$ & +0.15$^{+0.44}_{-0.44}$ & --- & --- & --- & --- & ---\vspace{2.0mm}\\
\hline
\end{tabular}
\label{Table:IndividualAbundances}
\end{table}
\end{landscape}

\begin{figure}
\includegraphics[scale=0.78,clip=]{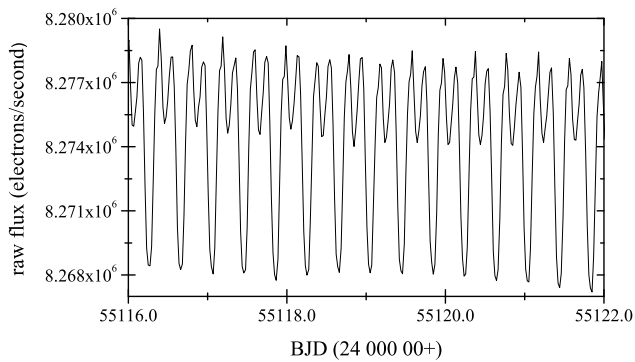}
\caption{{\small Part of the {\it Kepler} light curve of
KIC\,08018827.}} \label{K08018827_LC}
\end{figure}

\begin{figure}
\includegraphics[scale=0.85,clip=]{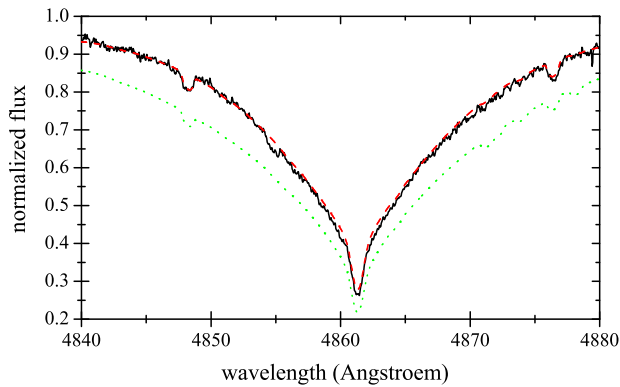}
\caption{{\small Same as Fig.~\ref{K02859567} but for
KIC\,07974841.}} \label{K07974841}
\end{figure}

\noindent KIC\,08018827 is reported by \citet{Balona2011b} to be of
B9 spectral type. Observed brightness variations are attributed to
rotation effects. Figure~\ref{K08018827_LC} shows a small portion of
the {\it Kepler} light curve of this star. The light curve seems to
be very regular showing both primary and secondary eclipses
occurring every $\sim$0.4 days. The fact that the secondary eclipse
occurs exactly at phase 0.5 relative to the primary minimum points
towards a circular orbit which is expected for such close binary
systems. Our two spectra is not enough to confirm binarity
spectroscopically, however. The spectroscopically derived value of
the effective temperature is right in between the two values
reported by \citet{Balona2011b} though closer to the one obtained
from the SED fitting. Both stars are found to be much hotter than is
expected from the KIC. In both cases we find that Fe and Si
abundances are consistent with the derived metallicity while Mg is
strongly enhanced for KIC\,08018827 and slightly depleted for
KIC\,07974841. The latter star additionally exhibits strong
depletion of He while nearly solar content of He is found for the
former. Figure~\ref{K07974841} compares the H$\beta$ line profile of
KIC\,07974841 (black, solid line) with the two synthetic profiles
computed from our optimised parameters (red, dashed line) and from
those listed in the KIC (green, dotted line). The difference is
clearly visible showing that the KIC underestimates the temperature.

\begin{figure}
\includegraphics[scale=0.75,clip=]{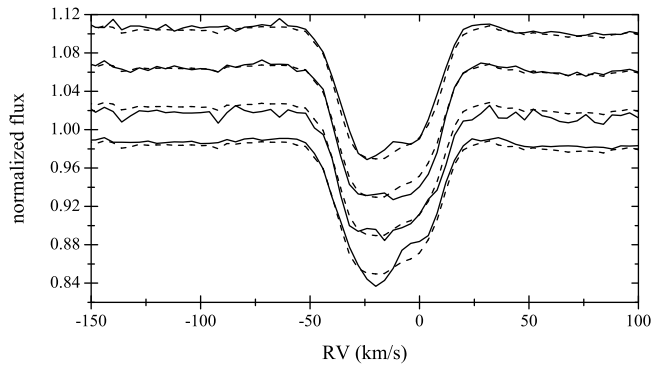}
\caption{{\small LSD-profiles computed from four individual spectra
of KIC\,08324268. The average, dashed profile is given for
comparison for better visibility of LPV.}} \label{K08324268}
\end{figure}

{\it KIC\,08324268:} The star was classified as a chemically
peculiar (CP) Si star by \citet{Zirin1951} and is found in the
Washington Double Star Catalog as WDS\,19586+4416.
\citet{Balona2011b} report an effective temperature of about
14\,400$\pm$2\,000~K which is much higher than the temperature we
derive in this work. From the light curve, the star seems to be a
monoperiodic variable with a period of about 2 days, attributed by
\citet{Balona2011b} to rotational modulation due to stellar surface
inhomogeneities. Our four spectra show clear variability caused by
moving bumps across the line profiles. Figure~\ref{K08324268} shows
these bumps in the least-squares deconvolved (LSD,
\citet{Donati1997}) profiles that are shifted in Y-axis for clarity.
Given that there is no signature of the second star in the  the
spectra nor Doppler shifts of the spectral lines pointing to
binarity, we can exclude binarity as a cause of the observed
variability. The strong variability of KIC\,08324268 does not allow
to obtain a reasonable fit of its mean spectrum. The derived
fundamental parameters of the star are strongly affected by the
variability and thus can be unreliable. Given that our fit of the
mean observed spectrum is fairly bad, no individual abundances are
available for this star.

{\it KIC\,12217324:} this is the sharpest-lined object among all
late B- early A-type stars in our sample. The derived effective
temperature is right in between the two values reported by
\citet{Balona2011b} and agrees with them within the error bars. Our
fundamental parameters are also in good agreement with those listed
in the KIC. The photometric variability of the star is attributed by
\citet{Balona2011b} to rotation effects. Our two spectra is not
enough to confirm these findings spectroscopically. The star does
not show any peculiarities in individual abundances except for a
small depletion of Mg and O and a slight excess of Cr.

\subsubsection{Position in the log(\te)--log(g) diagram}\label{Section:
hot_stars_HR_diagram}

\begin{figure}
\includegraphics[scale=0.85,clip=]{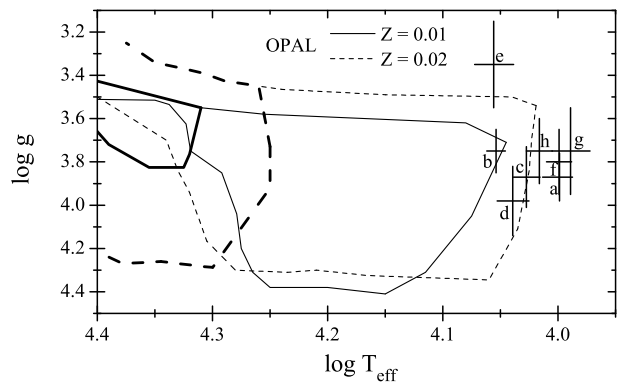}
\caption{{\small Location of the late B- early A-type stars (see
Table~\ref{Table: Hot_stars_classification} for labels) and the SPB
(thin lines) and $\beta$~Cep (thick lines) theoretical instability
strips in the log(\te)--log(g) diagram. The instability regions are
computed for two different metallicity values and are based on
\citet{Miglio2007}.}} \label{IS_hot_stars}
\end{figure}

In this section, we discuss the positions of the stars in the
log(\te)--log(g) diagram with respect to the theoretical SPB and
$\beta$\,Cep instability strips as described by \citet{Miglio2007}.
Figure~\ref{IS_hot_stars} shows the position of the stars in the
log(\te)--log(g) diagram together with the theoretical SPB and
$\beta$\,Cep instability strips, shown for two different values of
metallicity -- solar (dashed lines) and sub-solar (solid lines).

Table~\ref{Table: Hot_stars_classification} summarizes the results
of classification. According to their position in the
log(\te)--log(g) diagram, three of the hottest stars in our sample
(labels b, c, and d) are found to be SPB-type variables, one is
classified as ``possibly SPB'' (label h), and four stars are either
to cool (labels a, f, and g) or too evolved (label e) to be SPBs.

Two out of the four potential pulsators,  KIC\,08018827 and 12217324
(labels d and h), are claimed to be non-pulsating stars by
\citet{Balona2011b} who attribute their photometric variability to
rotation effects. KIC\,08018827 exhibits the light curve and Fourier
spectrum typical of close binary systems. Binarity is expected to
have an effect on the spectroscopically derived \te\ and \logg, and
thus on the position of the star in the log(\te)--log(g) diagram.
This makes spectroscopic classification of this star rather
uncertain. KIC\,12217324 is found close to the cool edge of the SPB
instability strip. Given that this star does not show remarkable LPV
nor peculiarities in terms of individual abundances, it is
considered by us as an SPB candidate pulsator.

Two out of three cool stars (labels f and g) are  low metallicity
stars. Moreover, for both stars the light curve morphology suggests
rotational modulation due to stellar surface inhomogeneities.
According to their position in the log(\te)--log(g) diagram, these
stars should not pulsate which confirms their photometric
classification. KIC\,02859567 (label a) is the only star among these
three that shows low-amplitude SPB-type pulsations in its Fourier
spectrum but is clearly outside the corresponding theoretical
instability region.

According to its position in the diagram, KIC\,08324268 (label e) is
too evolved to be an SPB variable. However, the LPV detected in the
spectra of this star is so strong that it prevents the derivation of
individual abundances and might affect the derived fundamental
parameters. Thus, similar to KIC\,08018827, the spectroscopic
classification of this star is rather uncertain.

\begin{table}
\tabcolsep 2.2mm \caption{\small Classification of late B- to early
A-type stars based on their position in the log(\te)--log(g)
diagram.}\label{Table: Hot_stars_classification}
\begin{tabular}{cclc}
\hline
KIC number\rule{0pt}{11pt} & \multicolumn{2}{c}{Variability} & label\\
                           & Diagram & \multicolumn{1}{c}{Other} &\\
\hline
03629496\rule{0pt}{9pt} & & beating star (SPB?)$^2$ & b\\
07974841\rule{0pt}{9pt} & SPBs & rotation+SPB$^1$ & c\\
08018827\rule{0pt}{9pt} & & rotation$^1$/binary$^3$ & d\\\cline{1-2}
12217324\rule{0pt}{11pt} & possibly SPB & rotation$^1$ &
h\\\cline{1-2}
02859567\rule{0pt}{11pt} & & possibly SPB$^3$ & a\\
08351193\rule{0pt}{9pt} & too cool & rotation$^1$ & f\\
10974032\rule{0pt}{9pt} & & rotation$^3$& g\\\cline{1-2}
08324268\rule{0pt}{11pt} & too evolved & rotation$^1$ & e\\
\hline\multicolumn{4}{l}{$^1$\citet{Balona2011b},
$^2$\citet{Balona2011a},\rule{0pt}{11pt}}
\\\multicolumn{4}{l}{$^3$this paper (light curve morphology, visual
inspection)}
\end{tabular}
\end{table}

\subsection{Intermediate A- to early F-type stars}\label{Section: Intermediate A- early F-type
stars}

 In this section, we discuss a sub-sample of 18 \GD/\DSct\
candidate stars. Twelve out of these 18 stars are divided into three
arbitrary groups, ``Stars with composite spectra'', ``Stars showing
clear LPV'', and ``Stars with peculiar abundances'', according to
their type of spectrum variability. Finally, we classify the stars
according to their positions in the log(\te)--log(g) diagram and
compare the results with the classification expected from the
analysis of $Kepler$ light curves.

\subsubsection{Stars with composite spectra}

\begin{figure}
\includegraphics[scale=0.72,clip=]{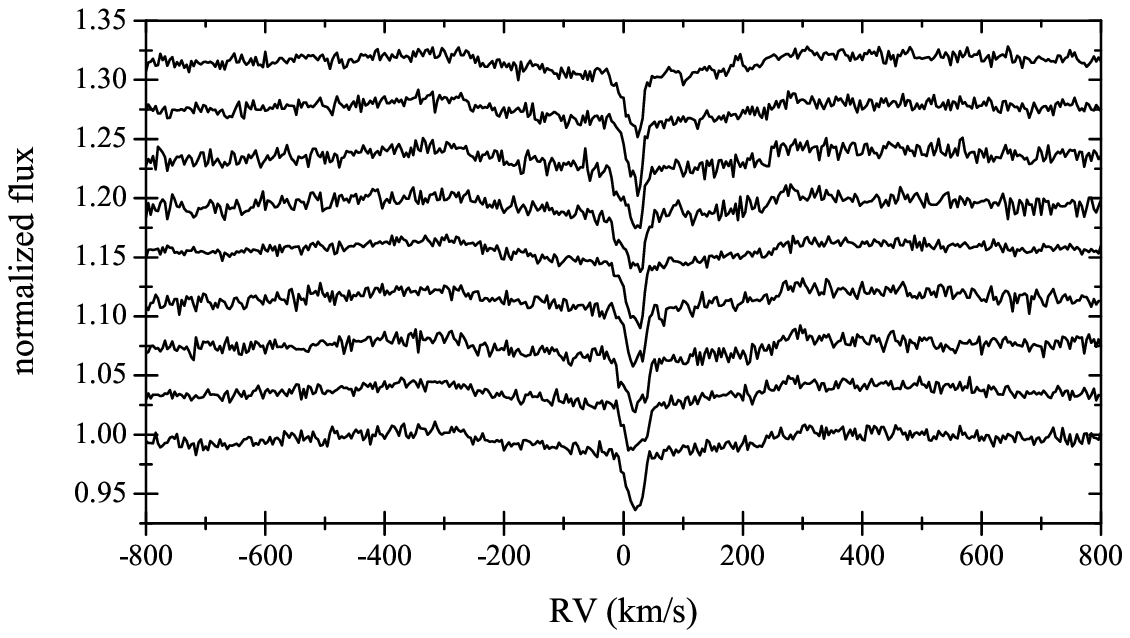}
\includegraphics[scale=0.72,clip=]{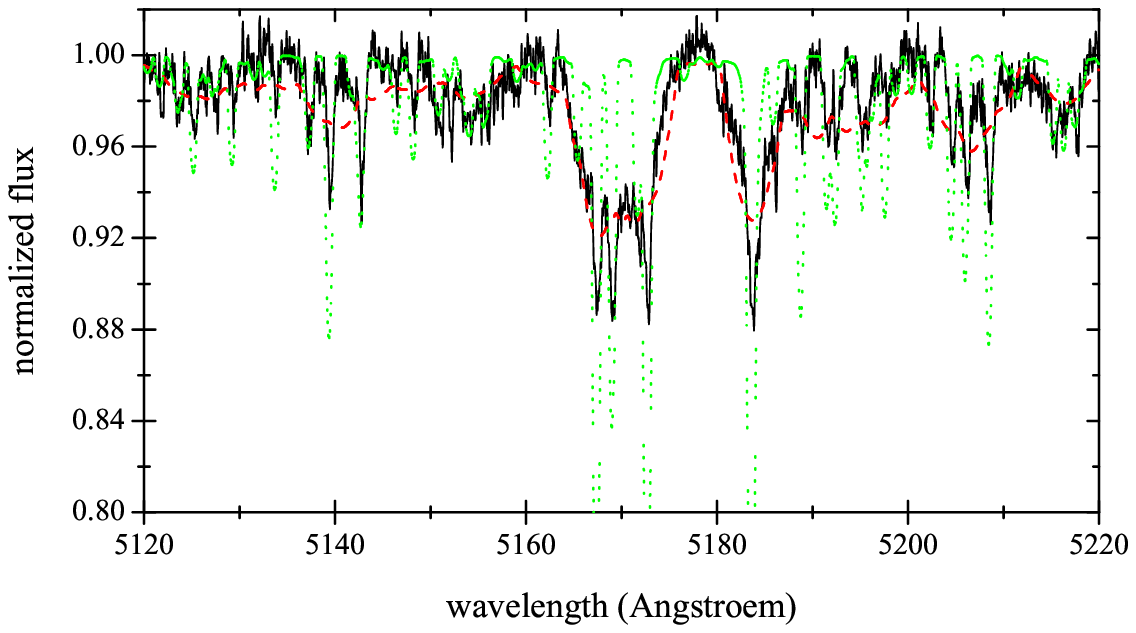}
\caption{{\small {\bf Top:} LSD-profiles computed from nine
individual spectra of KIC\,04180199. {\bf Bottom:} Fit of the
observed (black, solid line) by the synthetic spectra computed from
our optimised parameters and assuming \vsini\ of 180\,\kms\ (red,
dashed line) and 30\,\kms\ (green, dotted line).}} \label{K04180199}
\end{figure}

{\it KIC\,04180199, 11572666, {\rm and} 11874676:} three rapidly
rotating stars exhibiting similar photometric and spectroscopic
characteristics. The spectra are characterized by sharp central
cores superimposed on the broad absorption features extending
typically from --300 to +300\,\kms. Figure~\ref{K04180199} (top)
shows LSD-profiles computed from nine individual spectra of
KIC\,04180199. The profiles are shifted in Y-axis for clarity. The
lines have broad wings and timely variable sharp central cores. The
bottom panel of Figure~\ref{K04180199} compares the observed
spectrum of KIC\,04180199 with two synthetic spectra computed from
our optimised parameters and convolved with different \vsini\ of
180~\kms\ and 30~\kms. The positions of the majority of the spectral
lines in the green spectrum coincide with the sharp features seen in
the observed spectrum. The fact that both sharp and broad components
are stationary in RV excludes binarity as a cause of the observed
spectrum variability.

\citet{Mantegazza1996} report the detection of the composite
spectrum of the \DSct\ star X~Caeli (HD\,32846). Its spectral lines
show a complex behaviour with sharp central cores superimposed on
broad absorption features. The narrow core absorption was found to
be stable in time both in shape and position, excluding stellar
pulsations or orbital motion as first sight causes. Moreover, the
radial velocity of the core was found to be comparable to that of
the stellar barycenter making it improbable that it comes from a
foreground star, for example. The authors thus suggest a model
consisting of a \DSct\ variable star exhibiting broad spectral lines
that has a circumstellar shell characterized by the narrow
absorption cores superimposed on the photospheric lines of the
central object. Our stars show similar behaviour with the exception
that the narrow absorption core seems to be variable on a short time
scale (cf. Figure~\ref{K04180199}, top). None of the stars is
referred as a binary in the literature. Instead, KIC\,04180199 and
11572666 have been classified by \citet{Uytterhoeven2011} as hybrid
\GD--\DSct\ and KIC\,11874676 as \DSct\ pulsators.

All three stars have low metallicities with clear enrichment of Si,
Mg, and possibly C in their atmospheres. The difference between the
derived and the KIC temperatures does not exceed 400~K in the worst
case of KIC\,11874676 while the deviations in \logg\ are more
remarkable and reach 0.6~dex in the case of KIC\,11572666. Given the
complexity of the spectra described above, both the fundamental
parameters and the individual abundances are uncertain.

\subsubsection{Stars showing clear LPV}

\begin{figure}
\includegraphics[scale=0.75,clip=]{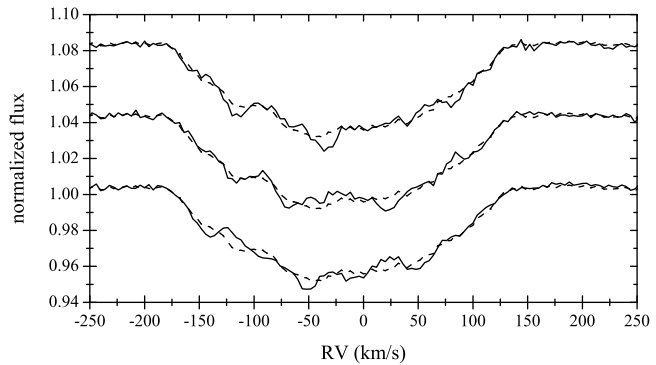}
\caption{{\small LSD-profiles computed from three individual spectra
of KIC\,02987660. The average, dashed profile is given for
comparison for better visibility of LPV.}} \label{K02987660_LSD}
\end{figure}

{\it KIC\,02987660, 07304385, 09351622 {\rm and} 10537907:} These
are four stars showing remarkable line profile variations in their
spectra. Figure~\ref{K02987660_LSD} shows individual LSD spectra of
KIC\,02987660 as an example of bumps moving across the profile,
possibly caused by stellar oscillations. Two of the stars,
KIC\,02987660 and 07304385, are reported by \citet{Uytterhoeven2011}
to be correspondingly \DSct\ and \GD-type pulsators, while the two
others, KIC\,09351622 and 10537907, are found to be of \GD-\DSct\
hybrid type. All four stars show individual abundances consistent
with the derived metallicity within the errors of measurement.

\subsubsection{Stars with peculiar abundances}\label{Section:
stars_peculiar_abundances}

{\it KIC\,05356349:} Though the star is in the Washington Double
Star Catalog (WDS J19196+4035AB), \citet{Uytterhoeven2011} classify
it as a single star of \GD-\DSct\ hybrid type. Our two spectra is
not enough to conclude on binarity. Some peculiarities in individual
abundances are detected, however. Ti and Cr are found to be
underabundant by correspondingly 0.6 and 0.5~dex compared to the
metallicity of the star, while Mg is enhanced by roughly the same
value.

{\it KIC\,07827131:} This star was first reported to be variable by
\citet{Magalashvili1976}. It is classified as \GD-\DSct\ hybrid
pulsator by \citet{Uytterhoeven2011}. According to our findings,
this is a low-metallicity star with unusually low Ti content
(--1.85~dex compared to the solar composition) and an enrichment of
Si of about 0.6~dex compared to the derived metallicity. As for the
previously discussed object, we have only two spectra which is not
enough to conclude on either LPV or binarity.

{\it KIC\,08489712, 08915335 {\rm and} 09291618:} These are three
stars with an enhanced Si content, up to 0.9~dex compared to the
derived metallicity in the case of KIC\,08915335. According to
\citet{Uytterhoeven2011}, two stars, KIC\,08489712 and 09291618, are
pure \GD\ and \DSct\ variables, accordingly, while the third star,
KIC\,08915335, has a light curve and Fourier spectrum characteristic
of \GD-\DSct\ type hybrid pulsators. None of the three stars shows
remarkable LPV nor Doppler shifts of spectral lines pointing to
binarity.

\subsubsection{Position in the log(\te)--log(g) diagram}

\begin{figure}
\includegraphics[scale=0.85,clip=]{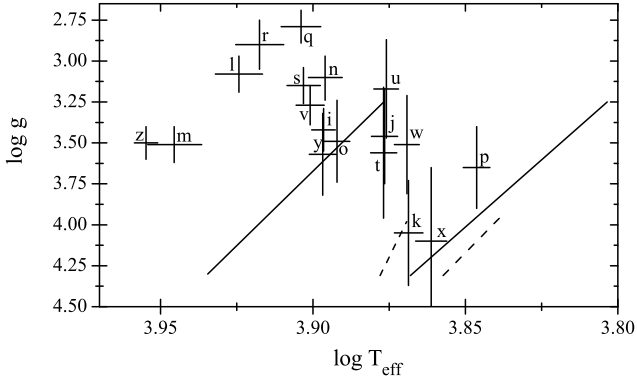}
\caption{{\small Location of intermediate A- to early F-type stars
(see Table~\ref{Table: Cool_stars_classification} for labels) and
the \GD\ (dashed lines) and \DSct\ (solid lines) observed
instability strips in the log(\te)-log(g) diagram.}}
\label{IS_cool_stars}
\end{figure}

Figure~\ref{IS_cool_stars} shows the positions of all intermediate
A- to early F-type stars of our sample in the log(\te)--log(g)
diagram, together with the \DSct\ (solid lines) and \GD\ (dashed
lines) observational instability strips as given by
\citet{Rodriguez2001} and \citet{Handler2002}, accordingly.

Table~\ref{Table: Cool_stars_classification} classifies the stars
according to their type of variability, listing the classifications
expected from their location in the log(\te)--log(g)
(Figure~\ref{IS_cool_stars}) and derived by \citet{Uytterhoeven2011}
from the frequency analysis of the $Kepler$ light curves. We find
three stars (labels m, v, and z) which are too hot and five stars
(l, n, q, r, and s) which are too evolved to be \GD\ or \DSct\
variables. Four of them, KIC\,05356349, 07827131, 08489712, and
08915335 (labels m, q, r, and s), exhibit peculiar individual
abundances and have been discussed in detail in the previous
section. KIC\,12153021 (label z) does not show any remarkable
variability in its light curve nor in the spectra and seems to be a
constant star. Three further stars, KIC\,04989900, 05437206, and
10096499 (labels l, n, and v) have been found by
\citet{Uytterhoeven2011} to show both \GD- and \DSct-type
pulsations, thus they appear to be hotter from our spectroscopic
analysis than it is expected from the photometric classification.

We confirm two \GD\ pulsators (labels k and x) and four \DSct\ stars
lying in the expected region of the log(\te)--log(g) diagram. Three
out of these six stars (labels k, x, and w) have been classified by
\citet{Uytterhoeven2011} as \GD-\DSct\ hybrid pulsators, two stars
(labels j, and t) were reported as pure \DSct\ variables, and one
(label p) as a \GD\ pulsator. Finally, four stars (labels i, o, u,
and y) are located close to the hot border of the \DSct\ instability
strip and are thus classified by us as ``possibly \DSct'' variables.
At least one of these stars (label u) is expected to be of lower
temperature from the theory of non-radial pulsations, however, as it
is reported by \citet{Uytterhoeven2011} to show \GD-like
oscillations in its light curve.

\section{Stellar Temperatures from Spectral Energy Distributions}

The effective temperature of the stars can be estimated from the
spectral energy distribution (SED). For our target stars, the SED
was constructed from literature photometry, using 2MASS
\citep{Sktrutskie2006}, Tycho $B$ and $V$ magnitudes
\citep{Hog1997}, USNO-B1 $R$ magnitudes \citep{Monet2003} and TASS
$I$ magnitudes \citep{Droege2006}, supplemented with CMC14 $r'$
magnitudes \citep{Evans2002} and TD-1 ultraviolet flux measurements
\citep{Carnochan1979}.

\begin{table}
\tabcolsep 3.0mm \caption{\small Classification of the intermediate
A- to early F-type stars based on their position in the
log(\te)--log(g) diagram.}\label{Table: Cool_stars_classification}
\begin{tabular}{cccc}
\hline
KIC number\rule{0pt}{11pt} & \multicolumn{2}{c}{Variability} & label\\
                           & Diagram & \multicolumn{1}{c}{Other} &\\
\hline
04180199\rule{0pt}{9pt} & & & k\\
11572666\rule{0pt}{9pt} & \raisebox{1.5ex}[-1.5ex]{\GD\ or hybrid} &
\raisebox{1.5ex}[-1.5ex]{hybrid$^1$} & x\\\hline
02987660\rule{0pt}{9pt} &  &  & j\\
09291618\rule{0pt}{9pt} & & \raisebox{1.5ex}[-1.5ex]{\DSct$^1$} &
t\\\cline{3-4} 07304385\rule{0pt}{9pt}&
\raisebox{1.5ex}[-1.5ex]{\DSct} & \GD$^1$& p\\\cline{3-4}
10537907\rule{0pt}{9pt} &  &  & w\\\cline{1-2}
09351622\rule{0pt}{9pt} &  & hybrid$^1$ & u\\\cline{3-4}
02571868\rule{0pt}{9pt}& & & i\\
06668729\rule{0pt}{9pt} & \raisebox{1.5ex}[-1.5ex]{possibly \DSct} & \DSct$^1$ & o\\
11874676\rule{0pt}{9pt} & &  & y\\\hline 10096499\rule{0pt}{9pt} & &
\GD$^1$ & v\\\cline{3-4} 05356349\rule{0pt}{9pt} & too hot &
hybrid$^1$ & m\\\cline{3-4} 12153021\rule{0pt}{9pt} & & not
pulsating$^2$ & z\\\hline
04989900\rule{0pt}{9pt} & & & l\\
05437206\rule{0pt}{9pt} & & & n\\
07827131\rule{0pt}{9pt} & too evolved &
\raisebox{1.5ex}[-1.5ex]{hybrid$^1$} & q\\
08915335\rule{0pt}{9pt} & &  & s\\\cline{3-4}
08489712\rule{0pt}{9pt} & & \GD$^1$ & r\\
\hline\multicolumn{4}{l}{$^1$\citet{Uytterhoeven2011}\rule{0pt}{11pt}}
\\\multicolumn{4}{l}{$^2$this paper (light curve morphology, visual
inspection)}
\end{tabular}
\end{table}

The SED can be significantly affected by interstellar reddening. We
have, therefore, estimated this effect from the equivalent widths of
the interstellar Na D lines present in our spectra. $E(B-V)$ was
calculated using the relation given by \cite{Munari1997}. For
resolved multi-component interstellar Na D lines, the equivalent
widths of the individual components were measured. The total
$E(B-V)$ in these cases is the sum of the reddening per component,
since interstellar reddening is additive \citep{Munari1997}. The SED
was de-reddened using the analytical extinction fits of
\cite{Seaton1979} for the ultraviolet and \cite{Howarth1983} for the
optical and infrared.

\te\ values were determined by fitting solar-composition
\cite{Kurucz1993} model fluxes to the de-reddened SED. For that, the
model fluxes were convolved with photometric filter response
functions. A weighted Levenberg-Marquardt non-linear least-squares
fitting procedure was used to find the solution that minimized the
difference between the observed and model fluxes. Since \logg\ is
poorly constrained by the SED, we fixed \logg=4.0 for all the fits.
The results are given in Table~\ref{Teff-SED}. The uncertainties in
\te\ include the formal least-squares error and adopted
uncertainties in $E(B-V)$ of $\pm$0.02 and \logg\ of $\pm$0.5 added
in quadrature.

\begin{table}
\caption{Stellar Effective Temperatures from Spectral Energy
Distributions}\label{Teff-SED}
\begin{tabular}{llll} \hline
KIC      & $E(B-V)$ &$T_{\rm eff}$    & SED Notes
\\\hline
02571868 & 0.01     &  8050 $\pm$ 330 & TD-1 \\
02859567 & 0.01     & 10070 $\pm$ 370 & TD-1 \\
02987660 & 0.01     &  7530 $\pm$ 300 & TD-1 \\
03629496 & 0.03     & 12130 $\pm$ 540 & TD-1 \\
04180199 & 0.08     &  7810 $\pm$ 370 & CMC14 $r'$ \\
04989900 & 0.06     &  8750 $\pm$ 230 & TD-1 \\
05356349 & 0.05     &  9150 $\pm$ 300 & TD-1 \\
05437206 & 0.04     &  7990 $\pm$ 320 & TD-1 \\
06668729 & 0.02     &  8120 $\pm$ 330 & TD-1 \\
07304385 & 0.06     &  7070 $\pm$ 320 &  \\
07827131 & 0.05     &  8580 $\pm$ 290 & TD-1 \\
07974841 & 0.07     & 10680 $\pm$ 420 & TD-1 \\
08018827 & 0.03     & 10240 $\pm$ 410 & TD-1 \\
08324268 & 0.05     & 10980 $\pm$ 520 & TD-1 \\
08351193 & 0.01     & 10760 $\pm$ 430 & TD-1 \\
08489712 & 0.08$^*$ &  9020 $\pm$ 330 & TD-1 \\
08915335 & 0.14$^*$ &  8600 $\pm$ 470 & CMC14 $r'$ \\
09291618 & 0.06     &  7760 $\pm$ 360 &  \\
09351622 & 0.06     &  7680 $\pm$ 340 &  \\
10096499 & 0.00     &  8370 $\pm$ 240 & TD-1 \\
10537907 & 0.06$^*$ &  7500 $\pm$ 340 & CMC14 $r'$ \\
10974032 & 0.02     &  9380 $\pm$ 290 &  \\
11572666 & 0.07$^*$ &  7260 $\pm$ 320 & CMC14 $r'$ \\
11874676 & 0.08     &  8300 $\pm$ 420 &  \\
12153021 & 0.01     &  8970 $\pm$ 310 & TD-1 \\
12217324 & 0.01     & 10360 $\pm$ 390 & TD-1 \\
\hline\multicolumn{4}{l}{$^*$indicates multi-component interstellar
Na D lines}
\end{tabular}
\end{table}

\begin{figure}
\centering
\includegraphics[scale=0.85,clip=]{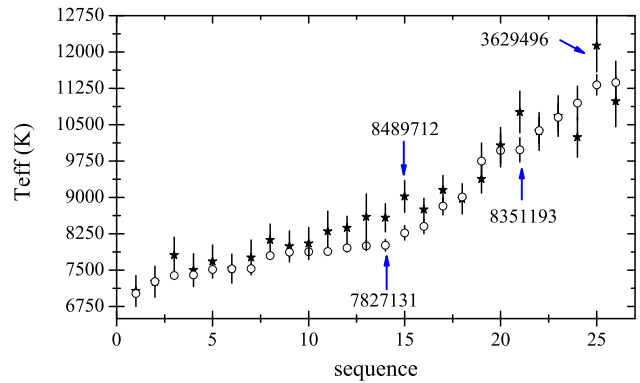}
\caption{{\small Comparison between the effective temperatures
derived spectroscopically (open circles) and from the SED fitting
(stars). The stars are sorted by the spectroscopic value starting
with the coolest object.}} \label{Comp_SED_teff}
\end{figure}

Figure~\ref{Comp_SED_teff} compares the spectroscopically derived
effective temperatures with those obtained from the SED fitting.
There are four stars (KIC\,03629496, 07827131, 08351193, and
08489712) for which the difference between the two temperatures is
larger than the quoted error bars, otherwise the values agree within
the error of measurement. For all four stars the SED temperatures
are higher than the spectroscopic value. Two stars, KIC\,07827131
and 08351193, are low metallicity objects both showing remarkable
enhancement of Si abundance. This peculiarity can possibly explain
the observed disagreement in temperatures. KIC\,08489712 also shows
peculiar behaviour in the sense of significant over- and
underabundances of Si and Ti, accordingly. There is no obvious
reason why the two temperatures differ by about 800~K for
KIC\,03629496, however.

\section{Comparison with the Kepler Input Catalog}\label{Section: Comparison with the
KIC}

\begin{figure}
\centering
\includegraphics[scale=0.85,clip=]{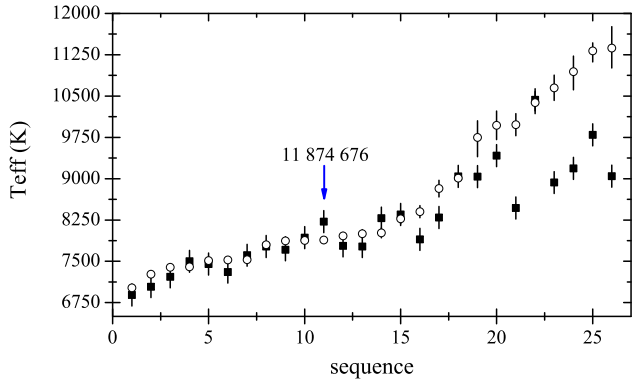}
\caption{{\small Same as Fig.~\ref{Comp_SED_teff} but for comparison
with the KIC values (filled boxes).}} \label{Comp_KIC_teff}
\end{figure}

\begin{figure}
\centering
\includegraphics[scale=0.87,clip=]{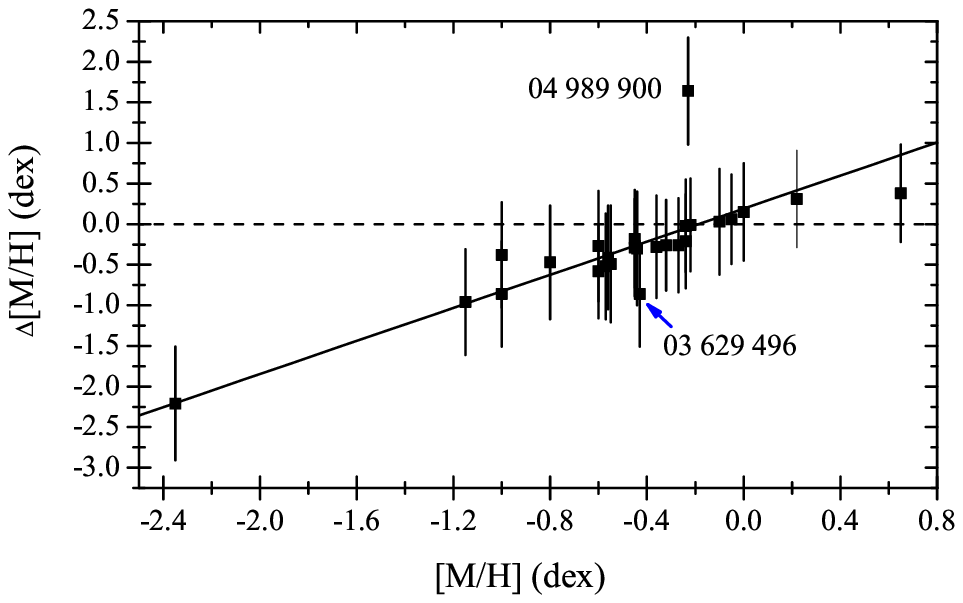}
\includegraphics[scale=0.87,clip=]{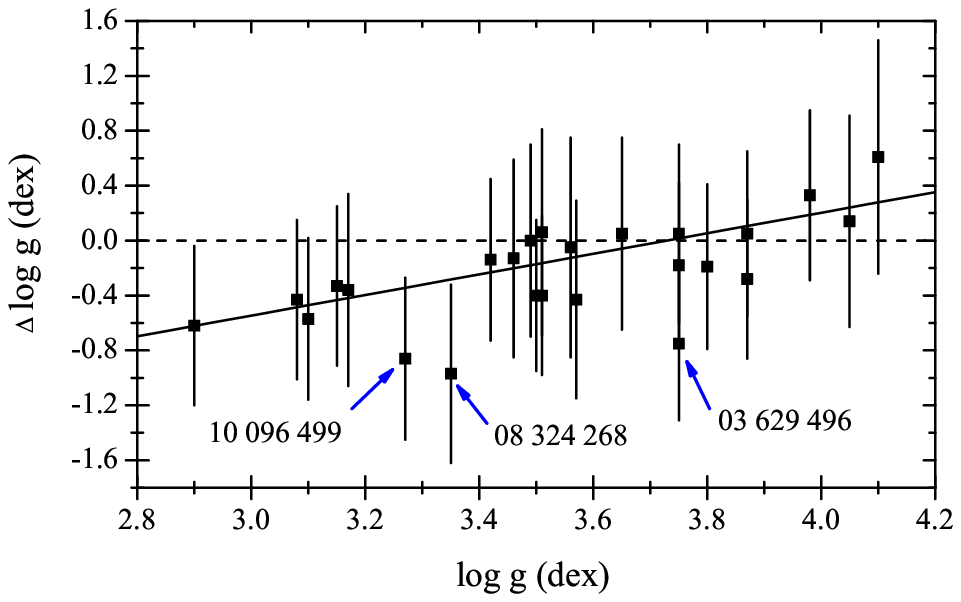}
\caption{{\small Comparison of the spectroscopically derived
metallicity [M/H] (top) and surface gravity \logg\ (bottom) with the
KIC values. The solid lines represent linear fits to the data points
after removing the marked outliers.}} \label{Comp_KIC_logg_metal}
\end{figure}

Figs.~\ref{Comp_KIC_teff} and \ref{Comp_KIC_logg_metal} compare the
spectroscopically derived atmospheric parameters (\te, \logg, and
[M/H]) with the KIC values. The typical errors of the KIC data of
$\pm$200~K for \te\ and of $\pm$0.5~dex for both \logg\ and
metallicity are assumed. As it was for the first time reported by
\cite{Molenda-Zakowicz2010} and later on confirmed by
\citet{Lehmann2011}, the KIC temperatures are systematically
underestimated for stars hotter than about 7\,000~K. The same
conclusion is valid for our stars: most of the stars from
intermediate A- to late B- spectral type have spectroscopic
temperatures remarkably higher than those listed in the KIC. The
stars of later spectral types and thus of lower temperatures show
good agreement with the KIC values within the error of measurements.
KIC\,11874676 seems to be the only star which does not follow this
general tendency as its spectroscopically derived \te\ value appears
to be by about 350~K lower than is given in the KIC (cf.
Figure~\ref{Comp_KIC_teff}). We think that this deviation comes from
the fact that KIC\,11874676 has a composite spectrum which lead to
wrong parameter values in our analysis.

\citet{Pinsonneault2012} published a revised temperature scale for
long-cadence targets in the KIC. The authors derive effective
temperatures based on griz filter photometry and compare their
temperature scale to the published infrared flux method scale for
V$_T$JK$_s$ \citep{Casagrande2010}. For field dwarfs,
\citet{Pinsonneault2012} find a mean shift toward hotter
temperatures relative to the KIC, of order 215 K, for both colour
systems. Unfortunately, only one of our targets is included in the
catalogue by \citet{Pinsonneault2012} and this and the seven of our
targets included in \citet{Casagrande2010} have temperatures in the
region where our values and the KIC values agree well.

The deviations between spectroscopic and KIC values in metallicity
and (to some extend) surface gravity seem to be correlated with the
parameter values itself. This is illustrated in
Fig.\,\ref{Comp_KIC_logg_metal} that shows the difference between
ours and the KIC values versus the spectroscopic values of the
parameters. The given error bars are the combined spectroscopic and
KIC errors. Stars that are outliers with respect to the observed
general tendency are marked by their KIC numbers and discussed in
the following.

{\it KIC\,03629496 {\rm and} 08324268:} These are the two hottest
stars in our sample both showing large discrepancy of more than
1\,500~K between spectroscopic and KIC temperatures. Thus, it is not
surprisingly at all that they also show large deviations of the
spectroscopically derived \logg\ from the KIC values. KIC\,03629496
additionally has larger $\Delta$[M/H] than it would be expected from
the observed general tendency for this parameter. In the case of
KIC\,08324268 neither our derived fundamental parameters nor the KIC
values provide satisfying fit of the observed spectrum, whereas our
parameters are much more appropriate than the KIC ones for
KIC\,03629496 (cf. the bottom panel of Figure~\ref{K02859567} for
the quality of the fit).

\begin{figure}
\includegraphics[scale=0.88,clip=]{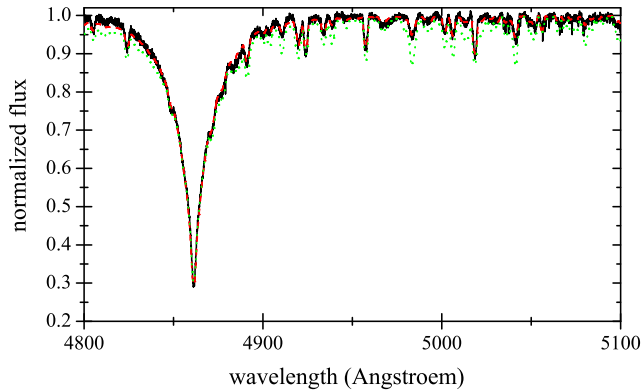}
\caption{{\small Fit of the observed spectrum of KIC\,10096499
(black, solid line) by synthetic spectra computed from our optimized
parameters (red, dashed line) and from the values given in the KIC
(green, dotted line).}} \label{K10096499}
\end{figure}

{\it KIC\,10096499:} The deviation of the optimised \logg\ from the
KIC value is larger than it would be expected from the general
tendency for this parameter. Our spectroscopic parameters provide a
much better fit of the observations than the KIC values.
Figure~\ref{K10096499} illustrates the quality of the spectrum fit
by two synthetic spectra computed from our optimised and KIC
parameters.

\begin{figure}
\includegraphics[scale=0.88,clip=]{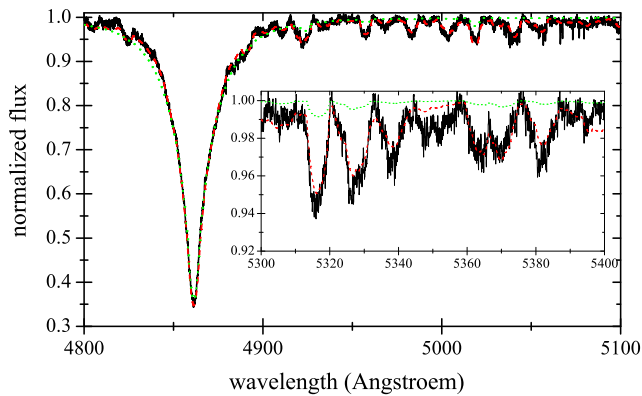}
\caption{{\small Same as Fig.~\ref{K10096499} but for
KIC\,04989900.}} \label{K04989900}
\end{figure}

{\it KIC\,04989900:} This star is an obvious outlier in the
[M/H]--$\Delta$[M/H] plane. The KIC metallicity of $-1.87$ is much
lower than our optimised value. Fig.~\ref{K04989900} compares the
observed (black, solid line) spectrum with two synthetic spectra
computed from our optimised (red, dashed line) and the KIC (green,
dotted line) fundamental parameters. It can be seen that the
KIC-based fit seriously underestimates the strengths of the metal
lines.

\section{Conclusions}\label{Section: Conclusions}

We used the spectral synthesis method to determine the fundamental
atmospheric parameters of 26 stars in the $Kepler$ satellite field
of view, of which 18 were proposed to be \GD/\DSct\ candidate
pulsators \citep{Uytterhoeven2011}, and compare our values with
those listed in the Kepler Input Catalog. Similar to the results
reported by \citet{Molenda-Zakowicz2010} and \citet{Lehmann2011}, we
find that the photometric KIC \te\ values are systematically
underestimated for stars hotter than about 7\,000~K. Deviations that
may reach several thousands Kelvin for hot stars \citep[see
e.g.,][]{Lehmann2011} are probably due to the interstellar reddening
that was not properly taken into account when estimating the KIC
temperatures.

Comparing the spectroscopically derived values of [M/H] and \logg\
with the KIC values, we find hints to a correlation of the
deviations with the parameter values itself. The correlation is
stronger for [M/H]. Four stars show larger deviations from this
general tendency. Except for KIC\,08324268 that shows strong LPV,
our parameters provide a much better fit of the observed spectrum
than the KIC values, however.

Comparison between the spectroscopically derived temperatures and
those obtained from the SED fitting reveals a good agreement for all
but four stars. For three stars, KIC\,07827131, 08351193, and
08489712, chemical composition peculiarities (particularly, Si and
Ti abundances) might be a possible reason for the observed
discrepancy.

The spectroscopically derived values of \te\ and \logg\ allow us to
place the stars in the log(\te)-log(g) diagram and classify them by
checking their location relative to the theoretical SPB/$\beta$\,Cep
(for B-type stars) and observational \GD/\DSct\ (for A- and F-type
stars) instability strips. Our spectroscopic classification was then
compared to the photometric one for both B-type
\citep{Balona2011a,Balona2011b} and A- and F-type
\citep{Uytterhoeven2011} stars. We confirm two SPB pulsators,
KIC\,03629496 and 07974841, and find two more stars, KIC\,08018827
and 12217324, that fall into or lie close the SPB instability region
but were classified by \citet{Balona2011b} as rotationally modulated
stars. We detected a clear signature of eclipses in the light curve
of KIC\,08018827. Binarity can have an impact on the
spectroscopically derived \te\ and \logg\ which in turn biases the
classification according to the type of variability. No such hints
including LPV have been found for KIC\,12217324. Three other stars,
KIC\,02859567, 08351193, and 10974032, are too cool to be SPBs and
too hot to be \DSct\ variables. According to the photometric
classification, however, one of these stars (KIC\,02859567) is
possibly a SPB pulsator while the variability of two others is
attributed to rotation effects. Finally, we find one star,
KIC\,08324268, which is too evolved to be a SPB pulsator but its
spectroscopic parameters might be biased by strong LPV detected in
the spectra.

Among the A-F type stars of our sample, we find two \GD\ pulsators
(KIC\,04180199 and 11572666) and four \DSct\ stars lying in the
expected region of the log(\te)--log(g) diagram. Two out of these
four stars show \DSct\ oscillations in their light curves, in one
star \DSct\ and \GD-typical oscillations co-exist, and one star is a
pure \GD\ pulsator. Four further stars, KIC\,02571868, 06668729,
09351622, and 11874676, are located at the hot border of the \DSct\
instability region. These stars are classified by us as ``possibly
\DSct'' pulsators and one of them surprisingly shows \GD-typical
oscillations in its light curves. Furthermore, we found eight stars
that are either too hot or too evolved to be \GD- or \DSct-type
variables, though all but one show characteristic \GD-like
pulsations in their light curves.

\citet{Uytterhoeven2011} presents a general characterization of 750
candidate A-F type stars in the $Kepler$ field of view and find
strong evidence for the existence of \GD\ and \DSct\ stars beyond
the edges of the current observational instability strips. The
authors conclude that a revision of the instability strips is needed
in order to accommodate the $Kepler$ \DSct\ and \GD\ stars. The same
conclusion is drawn by \citet{Grigahcene2010} concerning theoretical
\GD\ and \DSct\ instability strips. The authors characterize a
sample of 234 stars showing \DSct\ and \GD\ frequencies in their
light curves and find that the boarders of theoretical instability
strips are not a good match for the observations. Similar to the
findings of these two groups, we found that \GD-typical oscillations
are much more common among the \DSct\ stars than it is predicted by
the corresponding instability strips and conclude that a revision of
the latter is essential.

\section*{Acknowledgments} KU acknowledges financial support by the Spanish Ministry of Economy Competitiveness (MINECO) grant AYA2010-17803. The research leading to these results received funding
from the European Research Council under the European Community's
Seventh Framework Programme (FP7/2007–2013)/ERC grant agreement
n$^{\circ}$227224 (PROSPERITY). The authors are grateful to the
referee for very useful comments that led to significant improvement
of the paper. This publication makes use of data products from the
Two Micron All Sky Survey, which is a joint project of the
University of Massachusetts and the Infrared Processing and Analysis
Center/California Institute of Technology, funded by the National
Aeronautics and Space Administration and the National Science
Foundation. This research has made use of the SIMBAD database,
operated at CDS, Strasbourg, France.

{}

\label{lastpage}


\begin{thebibliography}{}
\bibitem[Aerts et al.(2010)]{Aerts2010} Aerts, C.,
Christensen-Dalsgaard, J., \& Kurtz, D.~W.\ 2010, Asteroseismology:
, Astronomy and Astrophysics Library, Volume .~ISBN
978-1-4020-5178-4.~Springer Science+Business Media B.V., 2010
\bibitem[Aihara et al.(2011)]{Aihara2011} Aihara, H., Allende
Prieto, C., An, D., et al.\ 2011, ApJS, 193, 29
\bibitem[Auvergne et al.(2009)]{Auvergne2009} Auvergne, M., Bodin, P., Boisnard,
L., et al.\ 2009, A\&A, 506, 411
\bibitem[Balona(2011)]{Balona2011a} Balona, L.~A.\ 2011, MNRAS,
415, 1691
\bibitem[Balona et al.(2011)]{Balona2011b} Balona, L.~A., Pigulski,
A., Cat, P.~D., et al.\ 2011, MNRAS, 413, 2403
\bibitem[Carnochan(1979)]{Carnochan1979}
Carnochan, D.J., 1979, Bull. Inf. CDS, 17, 78
\bibitem[Casagrande et al.(2010)]{Casagrande2010} Casagrande, L., Ram{\'{\i}}rez, I.,
Mel{\'e}ndez, J., Bessell, M., \& Asplund, M.\ 2010, A\&A, 512, A54
\bibitem[De Cat et al.(2004)]{DeCat2004} De Cat, P.,
Daszy{\'n}ska-Daszkiewicz, J., Briquet, M., Dupret, M.-A.,
Scuflaire, R., De Ridder, J., Niemczura, E., Aerts, C.\ 2004, IAU
Colloq.~193: Variable Stars in the Local Group, 310, 195
\bibitem[Donati et al.\ (1997)]{Donati1997} Donati, J.-F., Semel,
M., Carter, B.~D., Rees, D.~E., \& Collier Cameron, A.\ 1997, MNRAS,
291, 658
\bibitem[Droege et al.(2006)]{Droege2006}
Droege, T.F., Richmond, M.W., Sallman, M., 2006, PASP, 118, 1666
\bibitem[Dupret et al.(2005)]{Dupret2005} Dupret, M.-A., Grigahc{\`e}ne, A.,
Garrido, R., Gabriel, M., \& Scuflaire, R.\ 2005, A\&A, 435, 927
\bibitem[Evans et al.(2002)]{Evans2002} Evans, D.W., Irwin, M.J.,
Helmer, L., 2002, A\&A, 395, 347
\bibitem[Fossati et al.(2007)]{Fossati2007} Fossati, L., Bagnulo, S., Monier,
R., et al.\ 2007, A\&A, 476, 911
\bibitem[Fossati et al.(2008)]{Fossati2008} Fossati, L., Bagnulo, S.,
Landstreet, J., et al.\ 2008, A\&A, 483, 891
\bibitem[Fossati et al.(2009)]{Fossati2009} Fossati, L., Ryabchikova, T.,
Bagnulo, S., et al.\ 2009, A\&A, 503, 945
\bibitem[Gilliland et al. (2010)]{Gilliland2010} Gilliland R.~L.,
Brown T.~M., Christensen-Dalsgaard J., et al., 2010, PASP, 122, 131
\bibitem[Grevesse et al.(2007)]{Grevesse2007} Grevesse, N., Asplund,
M., \& Sauval, A.~J.\ 2007, SSRv, 130, 105
\bibitem[Grigahc{\`e}ne et al.(2010)]{Grigahcene2010} Grigahc{\`e}ne,
A., Antoci, V., Balona, L., et al.\ 2010, ApJL, 713, L192
\bibitem[Handler \& Shobbrook(2002)]{Handler2002} Handler, G., \& Shobbrook,
R.~R.\ 2002, MNRAS, 333, 251
\bibitem[H{\o}g et al.(1997)]{Hog1997}
H{\o}g, E., B{\"a}ssgen, G., Bastian, U., et al. 1997, A\&A, 323, 57
\bibitem[Howarth(1983)]{Howarth1983} Howarth, I.D., 1983, MNRAS, 203, 301
\bibitem[Kaye et al.(1999)]{Kaye1999} Kaye, A.~B., Handler, G.,
Krisciunas, K., Poretti, E., \& Zerbi, F.~M.\ 1999, PASP, 111, 840
\bibitem[Kupka et al.(2000)]{Kupka2000} Kupka, F.~G.,
Ryabchikova, T.~A., Piskunov, N.~E., Stempels, H.~C., \& Weiss,
W.~W.\ 2000, Baltic Astronomy, 9, 590
\bibitem[Kurucz(1993)]{Kurucz1993} Kurucz, R.L., 1993, Kurucz CD-ROM 13.
SAO, Cambridge, USA.
\bibitem[Lehmann et al.(2011)]{Lehmann2011} Lehmann, H., Tkachenko, A.,
Semaan, T., et al.\ 2011, A\&A, 526, A124
\bibitem[Magalashvili \& Kumsishvili(1976)]{Magalashvili1976} Magalashvili, N.~L., \&
Kumsishvili, J.~I.\ 1976, Information Bulletin on Variable Stars,
1167, 1
\bibitem[Mantegazza \& Poretti(1996)]{Mantegazza1996} Mantegazza, L., \& Poretti,
E.\ 1996, A\&A, 312, 855
\bibitem[Miglio et al.(2007)]{Miglio2007} Miglio, A.,
Montalb{\'a}n, J., \& Dupret, M.-A.\ 2007, Communications in
Asteroseismology, 151, 48
\bibitem[Molenda-Zakowicz et al.(2010)]{Molenda-Zakowicz2010}
Molenda-Zakowicz, J., Jerzykiewicz, M., Frasca, A., Catanzaro, G.,
Kopacki, G., \& Latham, D.~W.\ 2010, arXiv:1005.0985
\bibitem[Molenda-{\.Z}akowicz et al.(2011)]{Molenda-Zakowicz2011}
Molenda-{\.Z}akowicz, J., Latham, D.~W., Catanzaro, G., Frasca, A.,
\& Quinn, S.~N.\ 2011, MNRAS, 412, 1210
\bibitem[Molenda-{\.Z}akowicz et al.(2013)]{Molenda-Zakowicz2013} Molenda-{\.Z}akowicz, J., Sousa, S.G, Frasca, A., et al., 2013,
MNRAS, submitted
\bibitem[Monet et al.(2003)]{Monet2003}
Monet, D.G., Levine, S.E., Casian, B., et al. 2003, AJ, 125, 984
\bibitem[Munari \& Zwitter(1997)]{Munari1997} Munari, U., Zwitter, T.,
1997, A\&A, 318, 26
\bibitem[Pamyatnykh(2002)]{Pamyathykh2002} Pamyatnykh, A.~A.\ 2002,
Communications in Asteroseismology, 142, 10
\bibitem[Pinsonneault et al.(2012)]{Pinsonneault2012} Pinsonneault,
M.~H., An, D., Molenda-{\.Z}akowicz, J., Chaplin, W.~J., Metcalfe,
T.~S., \& Bruntt, H.\ 2012, ApJS, 199, 30
\bibitem[Rodr{\'{\i}}guez \& Breger(2001)]{Rodriguez2001} Rodr{\'{\i}}guez, E., \&
Breger, M.\ 2001, A\&A, 366, 178
\bibitem[Schmidt-Kaler(1982)]{Schmidt-Kaler1982} Schmidt-Kaler, Th. 1982, in Landolt-B\"{o}rnstein, ed. K.
Schaifers, \& H. H. Voigt (Springer-Verlag), 2b
\bibitem[Seaton(1979)]{Seaton1979} Seaton, M.J., 1979, MNRAS, 187, 73
\bibitem[Shulyak et al.(2004)]{Shulyak2004} Shulyak, D., Tsymbal, V.,
Ryabchikova, T., St{\"u}tz, C., \& Weiss, W.~W.\ 2004, A\&A, 428,
993
\bibitem[Sktrutskie et al.(2006)]{Sktrutskie2006} Skrutskie, M.F., Cutri,
R.M., Stiening, R., et al., 2006, AJ, 131, 1163
\bibitem[Tkachenko et al.(2012)]{Tkachenko2012} Tkachenko, A.,
Lehmann, H., Smalley, B., Debosscher, J., \& Aerts, C.\ 2012, MNRAS,
422, 2960
\bibitem[Tsymbal(1996)]{Tsymbal1996} Tsymbal, V.\ 1996, M.A.S.S.,
Model Atmospheres and Spectrum Synthesis, 108, 198
\bibitem[Uytterhoeven et al.(2010)]{Uytterhoeven2010a} Uytterhoeven, K.,
Szabo, R., Southworth, J., et al.\ 2010a, Astronomische Nachrichten,
331, P30 (arXiv:1003.6089)
\bibitem[Uytterhoeven et al.(2010)]{Uytterhoeven2010b} Uytterhoeven, K.,
Briquet, M., Bruntt, H., et al.\ 2010b, Astronomische Nachrichten,
331, 993
\bibitem[Uytterhoeven et al.(2011)]{Uytterhoeven2011} Uytterhoeven, K., Moya, A.,
Grigahc{\`e}ne, A., et al.\ 2011, A\&A, 534, A125
\bibitem[Waelkens(1991)]{Waelkens1991} Waelkens, C.\ 1991, A\&A, 246, 453
\bibitem[Zirin(1951)]{Zirin1951} Zirin, H.\ 1951, Harvard College
Obser. Bull., 920, 38
\end{thebibliography}
\end{document}